\documentclass[useAMS,usenatbib,usegraphicx]{mn2e}
\usepackage{times}

\def\xmm{{\em XMM-Newton}}
\def\c{{\em Chandra}}

\def\swift{{\em Swift}}

\def\wise{{\em WISE}}

\def\suzaku{{\em Suzaku}}
\def\nustar{{\em NuSTAR}}
\def\astroh{{\em ASTRO-H}}

\def\iras{{\em IRAS}}

\def\p{$\pm$}
\def\ltsim{\mathrel{\hbox{\rlap{\hbox{\lower4pt\hbox{$\sim$}}}\hbox{$<$}}}}
\def\gtsim{\mathrel{\hbox{\rlap{\hbox{\lower4pt\hbox{$\sim$}}}\hbox{$>$}}}}

\def\Msunpyr{M$_{\odot}\,$yr$^{-1}$}
\def\Msun{M$_{\odot}$}
\def\Lsun{L$_{\odot}$}
\def\micron{$\mu$m}
\def\cgs{erg\,s$^{-1}$\,cm$^{-2}$}

\def\ha{H$\alpha$}
\def\hb{H$\beta$}

\def\oiii{[O\,{\sc iii}]}

\def\nii{[N\,{\sc ii}]}

\def\l{$\lambda$}

\def\nev{[Ne\,{\sc v}]}

\def\nh{$N_{\rm H}$}

\def\eso565{ESO~565--G019}
\def\ftools{{\sc ftools}}
\def\xspec{{\sc xspec}}
\def\apec{{\sc apec}}
\def\pexrav{{\sc pexrav}}
\def\mytorus{{\sc mytorus}}
\def\torus{{\sc torus}}
\def\fscatt{$f_{\rm scatt}$}

\def\obj{NGC\,4785}
\def\thetainc{$\theta_{\rm inc}$}
\def\thetator{$\theta_{\rm tor}$}
\def\const{$C_{\rm XIS}^{\rm BAT}$}

\title[Suzaku observations of NGC 4785]{A Compton thick AGN in the barred spiral galaxy NGC 4785}

\author[P. Gandhi et al.]{P. Gandhi,$^1$ S. Yamada,$^{2}$ C. Ricci,$^3$ D. Asmus,$^4$ R.F. Mushotzky,$^5$ Y. Ueda,$^3$ Y. Terashima,$^6$\newauthor
V. La Parola$^7$\\
$^1$Department of Physics, Durham University, South Road, Durham DH1 3LE, UK.\\
$^2$Tokyo Metropolitan University, 1--1 Minami-Osawa, Hachioji-shi, Tokyo 192-0397, Japan\\
$^3$Department of Astronomy, Kyoto University, Kitashirakawa-Oiwake-cho, Sakyo-ku, Kyoto 606-8502, Japan\\
$^4$European Southern Observatory, Alonso de Cordova, Vitacura, Casilla 19001, Santiago, Chile\\
$^5$Department of Astronomy, University of Maryland, College Park, MD 20742, USA\\
$^6$Department of Physics, Ehime University, 2-5, Bunkyo-cho, Matsuyama, Ehime 790-8577, Japan\\
$^7$INAF, Istituto di Astrofisica Spaziale e Fisica Cosmica, via U. La Malfa 
153, 90146 Palermo, Italy
}

\begin{document}

\date{13 Aug 2014.}

\maketitle
\label{firstpage}

\begin{abstract}
We present X-ray observations of the active galactic nucleus (AGN) in \obj. The source is a local Seyfert 2 which has not been studied so far in much detail. It was recently detected with high significance in the 15--60\,keV band in the 66\,month \swift/BAT all sky survey, but there have been no prior pointed X-ray observations of this object. With \suzaku, we clearly detect the source below 10 keV, and find it to have a flat continuum and prominent neutral iron fluorescence line with equivalent width $\gtsim$\,1\,keV. Fitting the broadband spectra with physical reflection models shows the source to be a bona fide Compton thick AGN with \nh\ of at least $2\times 10^{24}$\,cm$^{-2}$ and absorption-corrected 2--10 keV X-ray power $L_{2-10}$\,$\sim$\,few times 10$^{42}$\,erg\,s$^{-1}$. Realistic uncertainties on $L_{2-10}$ computed from the joint confidence interval on the intrinsic power law continuum photon index and normalization are at least a factor of 10. The local bona fide Compton thick AGN population is highly heterogeneous in terms of \wise\ mid-infrared source colours, and the nucleus of \obj\ appears especially sub-dominant in the mid-infrared when comparing to other Compton thick AGN. Such sources would not be easily found using mid-infrared selection alone. The extent of host galaxy extinction to the nucleus is not clear, though \obj\ shows a complex core with a double bar and inner disk, adding to the list of known Compton thick AGN in barred host galaxies.
\end{abstract}

\section{Introduction}

Accurate assessment of the number density of highly obscured AGN remains a topic of intense interest. Compton thick AGN (with line-of-sight column densities of \nh\,$\gtsim$1.5\,$\times$\,$10^{24}$\,cm$^{-2}$) are expected to constitute a substantial fraction of the entire AGN population \citep{matt00, comastri95, fi, g03, gilli07, treister09, draperballantyne10, akylas12, ueda14}. Yet, their census appears to be far from complete. This is because photoelectric absorption and Compton scattering in the obscuring material (generally attributed to the $\sim$\,pc--scale circumnuclear torus of AGN unification schemes) results in severe attenuation of the direct AGN continuum below 10\,keV, and higher energy observations are needed for probing this component. At much higher columns, the only components to remain visible may be the reflected and scattered components, whose emitted flux is typically just a few per cent, or less, of the intrinsic AGN power \citep{iwasawa97, g14, arevalo14, balokovic14}. This makes identification and characterisation of Compton thick AGN (hereafter, CTAGN) a daunting task. 

Somewhat paradoxically, CTAGN at high redshift may be more easily identified because of the redshifting of high energy rest frame X-rays to below 10 keV. The exquisite point spread function and sensitivity of \c\ and deep follow up with \xmm\ over this energy range have resulted in the compilation of a few representative samples of distant CTAGN \citep{brightman14, georgantopoulos13}. 

In the local universe, probing the intrinsic continuum of CTAGN requires an ability to detect photons over the observed frame energy range above 10\,keV, where, until recently, sensitive instruments have been lacking. As a result, only about 20 CTAGN are known that are well characterized based upon a detection above 10\,keV as well as a fluorescence iron K$\alpha$ emission line at 6.4\,keV characteristic of reflection \citep{dellaceca08, goulding12, g14}. Increasing the number of these \lq bona fide\rq\ CTAGN is important for an accurate census of obscured accretion in the local universe. 

\swift/BAT \citep{swift, bat} is carrying out the most sensitive survey of the high energy sky, over a full observed energy range of 14--195\,keV. There are two complementary and independent processings of the BAT maps, one by the Goddard team \citep[cf. ][]{tueller08, tueller10, baumgartner13} and one by the Palermo team which uses modified energy bands \citep[cf. ][]{cusumano10_39, cusumano10_54, segreto10}. These have slightly differing optimizations, but there is broad agreement between both. The latest catalogs reach 4.8\,$\sigma$ flux limits of $\sim$\,10$^{-11}$\,erg\,s$^{-1}$\,cm$^{-2}$ over 50\,\%\ of the sky.\footnote{The base band for fluxes used by the Goddard group is 14--195 keV, while the Palermo group use 15--150 keV.}

Follow-up and cross-matching of this survey has provided new insights on AGN in hard X-rays. Results include a new hard X-ray luminosity function \citep{burlon11}, measuring the anti-correlation between luminosity and column densities in Compton-thin AGN \citep{winter09, vasudevan13_58m}, and insights into AGN triggering as a result of mergers \citep{koss10}. Cross-matching with all-sky infrared catalogs has provided new diagnostics for identification of various AGN classes \citep[e.g.][]{matsuta12, ichikawa12, maselli13}. Finally, follow-up with low energy X-ray missions, especially \suzaku, has proven to be very effective in identifying previously missed sources, including AGN with tori providing strong reflection and/or Compton thick absorption \citep{ueda07, severgnini11, g13_eso565}.

Here, we present new \suzaku\ follow-up observations of \obj, selected as a little studied Seyfert (Sy)\,2 galaxy first detected in hard X-rays on the 66\,month BAT maps by the Palermo group. The host galaxy is classified as a barred spiral [(R')SB(r)b] according to the Third Reference Catalog of Bright Galaxies \citep{rc3}. Despite having been recognized as a nearby ($z$\,=\,0.0123) Sy\,2 almost three decades ago \citep{fairall86}, the AGN properties remain unknown to a large degree. For instance, there has been no pointed observation of the source with any X-ray mission to date. And though the \oiii/\hb\ optical emission line flux ratio is known to be very high \citep{fairall86}, no absolute emission line fluxes have been published, to our knowledge.\footnote{Similarly, \nii\,\l 6584/\ha\ is also high in data presented by \citet{marquez04}.}

\suzaku\ reveals the source to be reflection-dominated over $\sim$\,2--10 keV. The combination of \suzaku\ and \swift/BAT data is used to model the broadband X-ray spectrum and show the source to be a bona fide CTAGN.   
We discuss the origin of the broadband X-ray emission, constraints on the intrinsic source power, and implications for multiwavelength studies of the CTAGN population in general. Luminosities quoted herein are based upon a redshift $z$\,=\,0.01311 corrected to the reference frame of the Cosmic Microwave Background and with a flat Planck cosmology with $H_{\rm 0}$\,=\,67.3\,km\,s$^{-1}$\,Mpc$^{-1}$ and $\Omega_\Lambda$\,=\,0.685 \citep{planckcosmology}, corresponding to a distance of 59\,Mpc. Note that a Tully-Fisher based distance of 49.6 Mpc is reported in the literature \citep{theureau07}, but other complementary redshift-independent measurements are lacking. So we use the redshift-based distance herein. All X-ray spectral fitting is carried out with the \xspec\ package v12.8.1g \citep{xspec} and uncertainties are quoted at 90\%\ confidence, unless stated otherwise.

\section{Observations}

\subsection{\swift}
\obj\ was first presented as a detection above 10 keV in the 66 month BAT catalog,\footnote{{\tt http://bat.ifc.inaf.it/}} where the X-ray detection has a signal-to-noise (S/N) of 7 in the 15--60\,keV band. More recent processings in progress have improved the S/N to 9.1 in the 100 month catalog.\footnote{Ibid.}. The X-ray source is offset by 0.54\,arcmin from the optical coordinates of \obj, well within the BAT positional uncertainty. 

Here, we use the most recent data from the Palermo team which is a cumulative exposure of $\approx$\,112 months. The response matrix is publicly available with the BAT 66 month catalog. 

\subsection{\suzaku}

\suzaku\ observed \obj\ on 2013-07-22 (ObsID 708003010) for an exposure of 79,388\,s. The low energy instrument called the X-ray Imaging Spectrometer (XIS; \citealt{xis}) and the high energy Hard X-ray Detector (HXD; \citealt{hxd, hxdinorbit}) were both operated in their standard modes with an XIS-nominal pointing. 

The standard \suzaku\ software available as part of \ftools\ v6.15.1 \citep{ftools} was used for data reduction. Cleaned event files were generated using standard filtering with pipeline v2.8.20.35 processing and used for the analysis herein. A 3.4\,arcmin radius circular aperture was used for integrating XIS source counts, and background counts were extracted from external source-free regions using a polygonal region. The \suzaku\ calibration database dated 2014-05-23 was used for generating response matrices (RMFs) and auxiliary response files (ARFs). The responses of the two front illuminated (FI) CCDs XIS0 and XIS3 match closely, so the spectra and responses were combined into one. The spectrum from the back illuminated (BI) detector XIS1 was treated separately. For fitting, the spectra were grouped to have a minimum S/N of at least 3 per energy bin after background subtraction. 

For HXD/PIN (sensitive between $\approx$\,15--60 keV), event extraction was carried out using the \ftools\ task {\tt hxdpinxbpi}. This extracts spectral counts, corrects for deadtime, and also returns a background spectrum which incorporates a typical contribution of the cosmic X-ray background (CXB) component to the \lq tuned\rq\ background model provided by the \suzaku\ team. Note that we used the latest version 2.2 (2.2ver1403) processing files which provide a more accurate estimate of the tuned non X-ray background for observations after August 2012. 

The source is too faint to be detectable in the higher energy HXD/GSO array, which is not considered here. 

\section{Results}

\subsection{\swift}
Fitting the BAT range with a power law (PL) returns a photon index of $\Gamma=1.67_{-0.36}^{+0.37}$, and a flux $F_{14-195}$\,=\,$1.2_{-0.3}^{+0.4}\times 10^{-11}$\,erg\,s$^{-1}$\,cm$^{-2}$ where the flux errors are determined using the \xspec\ {\tt cflux} command. 

\subsection{\suzaku}
The source is undetected in the HXD/PIN. The total PIN count rate (source+background) is $0.247\pm 0.002$ (15--60 keV), corresponding to a flux of $2.4\times 10^{-10}$\,\cgs\ when using a simple PL to characterise the HXD band. The typical systematic uncertainty on the NXB reproducibility is about 3\,\%\ for long observations \citep{fukazawa09} which results in an HXD/PIN flux upper limit of $F_{15-60}\sim 7.3\times 10^{-12}$\,\cgs. This limit is consistent with the \swift/BAT fitted model, which has a flux $F_{15-60}$\,=\,$5.3\times 10^{-12}$\,\cgs, and the PIN data are not included in the analysis hereafter. 

\begin{figure*}
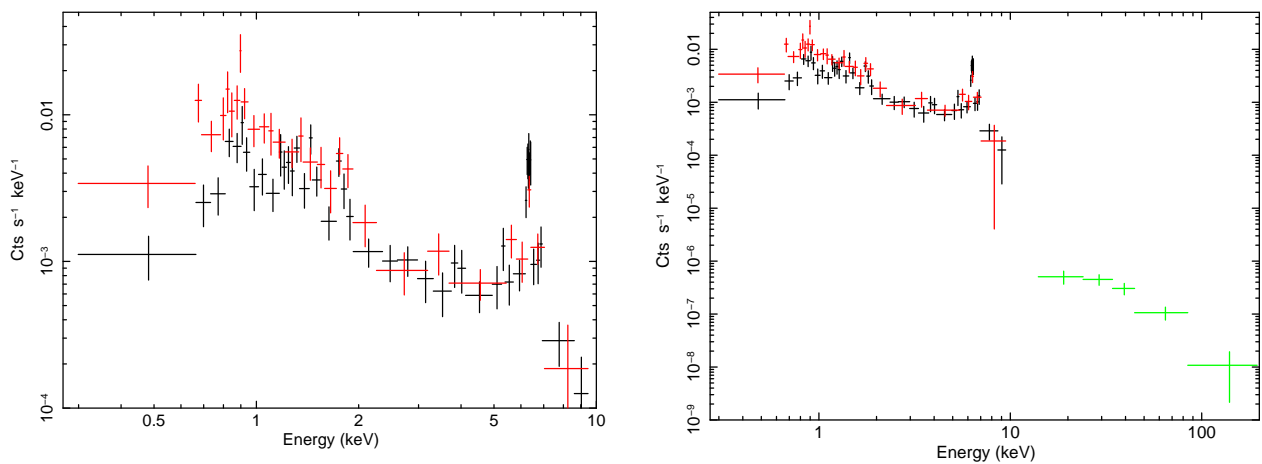

  \begin{center}
    \includegraphics[angle=270,width=8.5cm]{f1a.ps}
    \includegraphics[angle=270,width=8.5cm]{f1b.ps}
\caption{\suzaku\ spectra of \obj\ for the XIS on the left. The combined FI spectrum is in black and the BI spectrum in red. The right-hand panel additionally shows the \swift/BAT spectrum in green.   
 \label{fig:basic}}
  \end{center}
\end{figure*}

The source is clearly detected in the XIS FI and BI detectors. The spectra are shown in Fig.\,\ref{fig:basic}. The net count rates over the full range of $\sim$\,0.3--10\,keV are $9.1(\pm 0.4)\times 10^{-3}$ cts\,s$^{-1}$ (FI) and $1.4(\pm 0.1)\times 10^{-2}$ cts\,s$^{-1}$ (BI), respectively.

\subsection{Broadband modelling}

Fitting a simple PL model shows excesses above and below 2\,keV suggesting the presence of at least two separate components. Over the 2--10\,keV range, we find that a hard PL with $\Gamma$\,=\,0.69\,\p\,0.37 and a 6.4\,keV Fe K$\alpha$ line at the source systemic redshift can characterize the spectral shape adequately with $\chi^2$\,=\,42.5 for 41 degrees of freedom (dof). The observed flux is $F_{2-10}$\,=\,$2.4_{-0.4}^{+0.3}\times 10^{-13}$\,erg\,s$^{-1}$\,cm$^{-2}$ and the Fe line equivalent width EW\,=\,$1.7_{-0.4}^{+0.7}$\,keV. Such a flat spectral shape and strong Fe line are characteristic of reflection-dominated spectra as may be expected in CTAGN, so we proceeded to fit several standard models of reflection.

\subsubsection{Reflection models}
As a first attempt, we used the \pexrav\ model \citep{pexrav} to fit the reflection continuum. \pexrav\ assumes a slab geometry with an infinite optical depth of the reflector. It does not model the corresponding fluorescence emission lines expected, so a Gaussian component with fixed rest-frame energy of 6.4\,keV and width $\sigma$\,=\,50\,eV was included to simulate the strongest expected neutral Fe K$\alpha$ line. A direct transmitted PL with normalization and photon index tied to those of \pexrav\ was included separately and absorbed by both photoelectric absorption and Compton scattering associated with the torus using the {\sc zphabs} and {\sc cabs} models in \xspec, respectively. The \pexrav\ reflection continuum was assumed not to be absorbed by the torus, as may be expected if the reflection arises in the visible portions of the back face of the torus, for example. Solar abundances were assumed. This model is referred to as \lq Model P\rq.

\pexrav\ has historically been a popular model for fitting reflection spectra, and it is useful for comparison with published analyses of other sources. But since it does not self-consistently include fluorescence emission and requires ad-hoc assumptions about the visible reflector solid angle, we also fitted more physically self-consistent reflection models \mytorus\ \citep[][hereafter, \lq Model M\rq]{mytorus} and \torus\ \citep[][hereafter \lq Model T\rq]{brightmannandra11}. These models both describe photoelectric absorption, Compton scattering and fluorescence iron emission from a toroidal gas structure centered on the AGN. There are differences between the two models which allow an investigation of the systematics associated with the unknown geometry of the obscurer. Whereas the geometry assumed by \mytorus\ is a doughnut shape, with a line-of-sight column density \nh(los) tied to the inclination angle (\thetainc) and the equatorial column \nh(eq), \torus\ takes a conical section of a sphere with \nh(los) being independent of \thetainc\ for any line-of-sight passing through the torus. \mytorus\ assumes a fixed covering factor of 0.5, whereas this is allowed to vary in \torus\ through a variable opening angle \thetator. \mytorus\ allows the direct component to be decoupled from the scattered ({\sc mytoruss}) and fluorescence ({\sc mytorusl}) components and investigated independently of each other, whereas \torus\ does not. Tabulated models are publicly available for both models up to column densities of \nh(eq)\,=\,$10^{25}$\,cm$^{-2}$ (for \mytorus) and \nh\,=\,$10^{26}$\,cm$^{-2}$ (for \torus), respectively. Both models assume Solar abundances. 

\subsubsection{Additional model components}
With an angular resolution of 2\,arcmin (half power diameter), \suzaku\ cannot resolve out X-rays emitted by sources other than the direct AGN emission and torus reflection. These sources include emission from extended gas photoionised by the AGN, thermal emission from a hot interstellar medium, intrinsic AGN emission that is scattered from reflecting media above the torus, and emission from other point sources such as X-ray binaries. These sources may all dominate at low energies, typically below $\sim 2$\,keV where the direct AGN emission is completely absorbed. Distinguishing between these various sources generally requires high spectral resolution and high sensitivity soft X-ray data \citep[e.g. ][]{guainazzi07}. In the absence of such data, we simply parametrised the soft X-rays regime using a thermal model with the \apec\ code \citep{apec} in conjunction with a scattered PL (with scattering fraction \fscatt\ relative to the intrinsic AGN continuum), and discuss its feasibility in the Discussion section. These were added to all three reflection models described above. Note that the spectral portions between energies of 1.7--1.9 keV and 2.1--2.3 keV are ignored because of instrumental calibration uncertainties related to the Si and Au edges. Finally, fixed Galactic absorption of $1.1\times 10^{21}$\,cm$^{-2}$ \citep{dickeylongman90} was also included in all spectral fits. 
\newline

\noindent
The \xspec\ notation used to describe the three models is as follows, with explanatory associations in square brackets:

\begin{eqnarray}
\textsc{model\ P\ =\  const $\times$ phabs[} \mapsto N_{\textrm {\tiny H}}^{\textrm {\tiny Gal}}{\textsc ]} \times (\ \textrm{\textsc{apec}} + \nonumber\\
\textsc{pow} * \textsc{cabs} * \textsc{zphabs[} \mapsto N_{\textrm {\tiny H}}(\rm los){\textsc ]} + \nonumber\\
 \textsc{pexrav} + \textsc{zgauss} + \nonumber \\
\textsc{const [} \mapsto f_{\rm scatt} {\textsc ]}\times\textsc{pow}\ ), \nonumber
\end{eqnarray}

\noindent
and

\begin{eqnarray}
\textsc{model\ T\ =\  const $\times$ phabs[} \mapsto N_{\textrm {\tiny H}}^{\textrm {\tiny Gal}}{\textsc ]} \times (\ \textrm{\textsc{apec}} + \nonumber\\
\textsc{pow} * \textsc{atable}\{{\tt torus1006.fits}\} + \nonumber\\
\textsc{const [} \mapsto f_{\rm scatt} {\textsc ]}\times\textsc{pow}\ ), \nonumber
\end{eqnarray}

\noindent
and

\begin{eqnarray}
\textsc{model\ M\ =\  const $\times$ phabs[} \mapsto N_{\textrm {\tiny H}}^{\textrm {\tiny Gal}}{\textsc ]} \times (\ \textrm{\textsc{apec}} + \nonumber\\
\textsc{pow} * \textsc{etable} \{{\tt mytorus\_Ezero\_v00.fits}\} + \nonumber\\
\textsc{atable}\{{\tt mytorus\_scatteredH500\_v00.fits}\} + \nonumber\\
\textsc{atable}\{{\tt mytl\_V000010nEp000H500\_v00.fits}\} + \nonumber\\
\textsc{const [} \mapsto f_{\rm scatt} {\textsc ]}\times\textsc{pow}\ ). \nonumber
\end{eqnarray}

\subsection{Combined \suzaku\ and \swift\ fits}
The \swift/BAT observation provides a long-term average of the source flux. The \suzaku\ HXD/PIN flux limit is consistent with the \swift/BAT detected flux over 15--60 keV, meaning that the source could not have brightened significantly during the \suzaku\ observation with respect to the long-term average. We cannot rule out source fading, though CTAGN are generally dominated by reflection and variability is mitigated on light travel times across the torus. We thus fitted the \suzaku\ and \swift\ data simultaneously, accounting for flux variability and cross-calibration uncertainties using a scalar ({\sc constant}) parameter \const\ in the fits. This is fixed to 1 for the XIS FI and allowed free to vary for XIS BI as well as BAT. A caveat to be kept in mind is that we cannot rule out a \lq changing-look\rq\ nature with significant spectral variability using this approach.  

\begin{figure*}
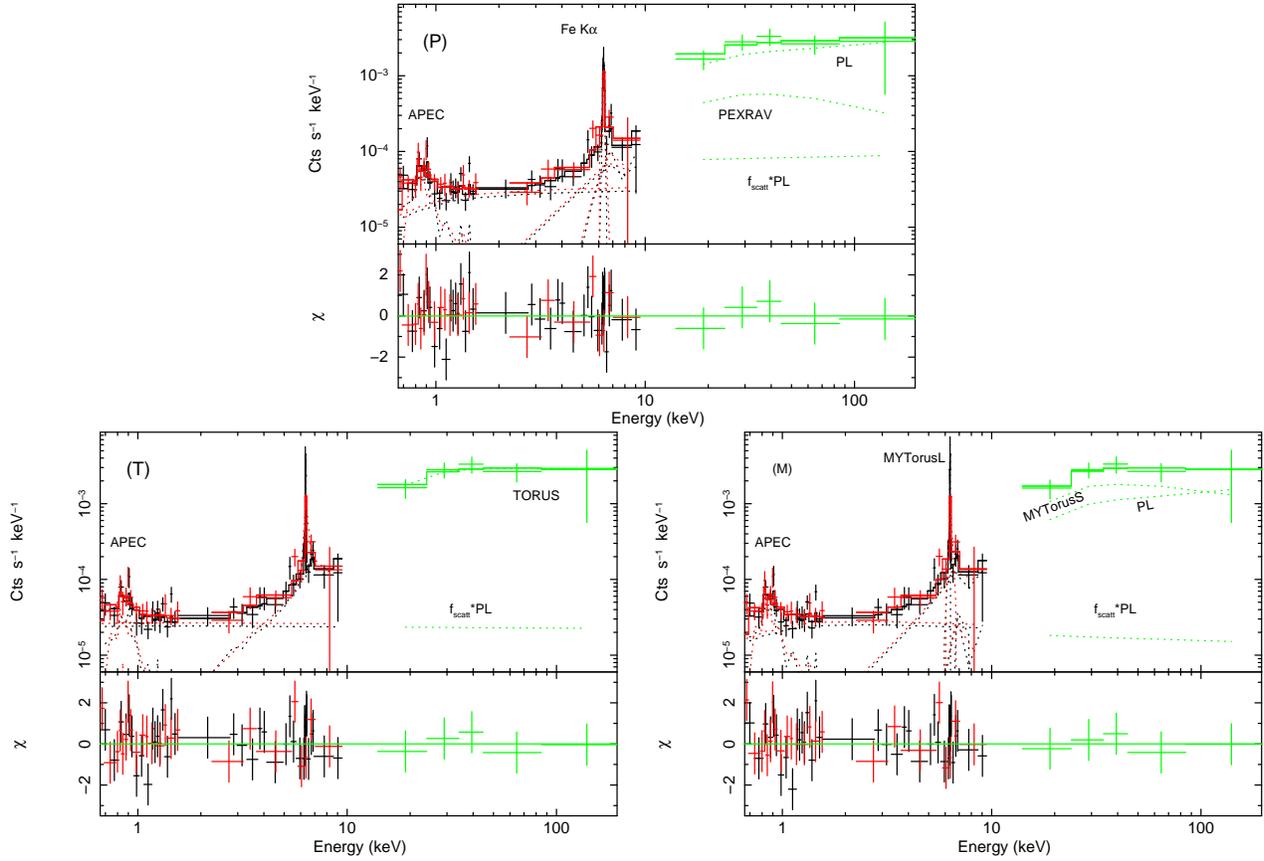

  \begin{center}
    \includegraphics[angle=270,width=8.5cm]{f2a.ps}
\vfill
\hfill
    \includegraphics[angle=270,width=8.5cm]{f2b.ps}
    \includegraphics[angle=270,width=8.5cm]{f2c.ps}
\caption{Broadband fits of models P (Top), T (Left) and M (Right) to the combined \suzaku\ and \swift\ spectrum of \obj. The panels show the unfolded models in $E F_E$ units. Residuals in units of sigmas with size one error bars are shown in the panels at the bottom. The \apec\ component dominates at the softest energies, and the scattered PL over $\approx$\,1--4\,keV. In models P and M, the contributions of the direct absorbed PL (peaking around 150\,keV), the Compton-scattered torus emission (peaking around 30 keV) and the fluorescence emission lines are separated. The \torus\ model (Model T) does not distinguish these contributions and only the total emission of the torus is plotted.  
 \label{fig:xspec}}
  \end{center}
\end{figure*}

Fig.\,\ref{fig:xspec} shows results of the fits for the three reflection models that we attempted, and the best fit parameters are listed in Table\,\ref{tab:x}. Model P yields an acceptable fit with a column density \nh(los)\,$\approx$\,1.4\,$\times$\,10$^{24}$\,cm$^{-2}$ that is borderline Compton thick. The data quality are not high enough to constrain \thetainc\ and the reflection fraction parameter $R$. The latter was fixed at $R$\,=\,--1 which is equivalent to a geometry where half of the intrinsic AGN emission is reflected, as may be expected for a sky covering factor of 0.5 for the torus. The fits prefer a near edge-on inclination model limit of cos(\thetainc)\,=\,0.05, though this is formally unconstrained. The fitted AGN photon index of $\Gamma$\,=\,1.9 is quite typical of bright well-studied local AGN (e.g. \citealt{mateos05_wide, piconcelli05}). The high Fe line EW(K$\alpha$)\,$\approx$\,1\,keV is fully consistent with a Compton thick absorbing column. However, there is a clear offset between the XIS and BAT with the cross-normalization constant of BAT relative to XIS being significantly larger than 1, and possibly much larger than 2. This could be caused by flux variability as the two datasets are not strictly simultaneous. But it could also be related to the limitations of the \pexrav\ model. Since this model does not self-consistently model the fluorescence emission strength, and since the transmitted and reflected components (likely dominating above and below 10 keV, respectively) are decoupled, there can be significant degeneracy when fitting non-simultaneous data covering non-overlapping energy ranges. In fact, as we show below, use of the more physical torus reflection models do not require large cross-calibration factors. 

We first considered the \torus\ model T. The fitted intrinsic continuum PL has a photon index $\Gamma$\,=\,2.1 and is absorbed by a high column density of \nh\,$>$\,$1.6\times 10^{24}$\,cm$^{-2}$ consistent with being Compton thick at 90\,\%\ confidence. The upper \nh\ limit is unconstrained, with the model threshold of \nh\,=\,10$^{26}$\,cm$^{-2}$ allowed. The Fe line is treated self-consistently in the \torus\ model as fluorescence from the torus. In order to gauge the line strength separately, we fitted an adhoc model with a power law continuum fixed to an approximation of the continuum fitted by the \torus\ model between energies of 5.5 and 7.0 keV (excluding the line), and then overlaid a Gaussian component to model the K$\alpha$ line. For the XIS FI data, this yielded EW(K$\alpha$)\,$\approx$\,0.8 keV. Allowing the continuum level to vary increases the line strength to EW(K$\alpha$)\,$\approx$\,1.0 keV.

The cross-normalization constants between the XIS FI and BI, and also between XIS FI and BAT, are both consistent with 1 within the uncertainties, though a factor of up to $\approx$\,2 variation between BAT and XIS cannot be excluded. \thetainc\ is unconstrained, so was fixed to its maximum value of 87.1\,deg as is common practice \citep{brightman14}. This allows the full range of \thetator\ values to be investigated. We find that \thetator\ values down to the lowest allowed model threshold of 26\,deg are allowed. But \thetator\,$>$72 deg are excluded. A scattered PL is required though \fscatt\ is constrained to be small below $\sim 1$\,\%. Note that we fixed the photon index of the scattered component to be the same as that of the intrinsic AGN PL. The temperature of the \apec\ component is $\sim 0.7$\,keV in all three models. 

We next tried replacing the torus reflection component with the simplest version of \mytorus\ (model M) which couples the normalizations and column densities of the scattered and fluorescence components to the zeroth order absorption distortion model of the direct continuum. The softer \apec\ and scattered components were included as before. As seen in Fig.\,\ref{fig:xspec} and Table\,\ref{tab:x}, this can yield a fit as good as Model T, again with a Compton thick obscurer solution with \nh(los)\,=\,$(2-7)\times 10^{24}$\,cm$^{-2}$. The opening angle of the torus is effectively fixed in the coupled \mytorus\ model, and a broad range of inclination angles \thetainc\,=\,64--85\,deg is allowed.

\begin{table*}
  \begin{center}
  \caption{Results of X-ray spectral fitting to NGC\,4785\label{tab:x}}
  \begin{tabular}{lccccr}
\hline
Component          &  Parameter       &     Model P         & Model T     &    Model M   & Units \\
\hline
\hline
\apec              & $kT$             & $0.68_{-0.10}^{+0.08}$ & $0.68_{-0.12}^{+0.14}$ & $0.68_{-0.02}^{+0.03}$ & keV\\
Absorber/Reflector & \nh(eq)          & --                  & --               & 2.9$_{-0.8}^{+4.1}$  &  $\times$ 10$^{24}$ cm$^{-2}$\\
                   & \nh(los)         & 1.4$_{-0.9}^{+0.5}$    & $2.2_{-0.6}^{+u}$  & 2.7$_{-0.8}^{+3.8}$  &  $\times$ 10$^{24}$ cm$^{-2}$\\
                   & $\theta_{\rm inc}$ & 87$_{-u}^{+u}$              & $87^f$          &  $79_{-15}^{+6}$   & deg\\
                   & $\theta_{\rm tor}$ & --                  & $57_{-32}^{+14}$  & --                & deg\\
                   & $R$              & --1$^f$             & -- & -- &  \\
                   & EW(Fe K$\alpha$) & 1.1$_{-0.4}^{+u}$     & $^c$ & $^c$  &  keV \\
AGN continuum      & $\Gamma$         & $2.0_{-0.4}^{+0.4}$    & $2.1_{-0.3}^{+0.4}$ & $2.1_{-0.4}^{+0.5}$ &  \\
Diffuse Scattering & \fscatt          & $14.8_{-9.9}^{+280}$              & $4.1_{-3.9}^{+8.8}$& $2.2_{-2.0}^{+7.5}$ &  $\times$ 10$^{-3}$\\
$C_{\rm XIS\ FI}^{\rm XIS\ BI}$ cross-calib & {\sc const}   & $1.11_{-0.16}^{+0.19}$ & $1.10_{-0.15}^{+0.17}$ & $1.10_{-0.16}^{+0.18}$ & \\
$C_{\rm XIS\ FI}^{\rm BAT}$ cross-calib    & {\sc const}   & $2.48_{-0.99}^{+5.32}$ & $1.07_{-0.41}^{+1.08}$ & $0.85_{-0.54}^{+1.11}$ & \\
                   &                  & & &  & \\
$\chi^2$/dof       &                  & 78.5/88 & 76.7/89 & 76.5/89 & \\
\hline
\end{tabular}~\par
$^u$unconstrained. $^f$fixed. $^c$line is produced self-consistently in the model.\\
Model P: \pexrav\ component fit \citep{pexrav}.\\
Model M: \mytorus\ coupled component fit \citep{mytorus}.\\
Model T: \torus\ model component fit \citep{brightmannandra11}.
\end{center}
\end{table*}

\section{Discussion}

We have presented the first pointed X-ray observations of \obj. The source has long been known as a relatively nearby Sy\,2, but little is known of the intrinsic AGN properties. With \suzaku, in combination with \swift/BAT, we show the source to be a bona fide CTAGN. This is based upon broadband X-ray spectral modelling over the range of $\approx$\,0.3--195 keV and the detection of a prominent neutral Fe K$\alpha$ emission line, both of which can be self-consistently fitted with the latest models incorporating Compton scattering off toroidal obscurers. 

\subsection{Broadband spectral components}
\label{sec:discussion1}

The obscuring column density along the line-of-sight is constrained to be greater than 2\,$\times$\,10$^{24}$\,cm$^{-2}$ for both the torus models that we tried, and near Compton thick in the \pexrav\ model. Although all three models are approximately consistent, the fits described in the previous section demonstrate that the physical torus models can effectively produce more spectral curvature than \pexrav, and can fit both the XIS and BAT regimes without the need for large cross-calibration factors. Furthermore, unlike \pexrav, these models self-consistently produce fluorescence emission. Therefore, for the remainder of this paper, we concentrate on the results of the torus models. 

Model M demands \nh\,$\approx$\,(2--7)\,$\times$\,10$^{24}$\,cm$^{-2}$. For model T, the \nh\ upper limit is apparently unconstrained with values as high as the model threshold of \nh\,=\,10$^{26}$\,cm$^{-2}$ allowed. But examining these solutions with extreme \nh\ values shows that they require a significant mismatch between the \suzaku\ and \swift\ flux levels, with the cross-normalization constant $C_{\rm XIS\ FI}^{\rm BAT}$ being $\approx$\,2. Although such strong variability cannot be completely ruled out, it is unlikely for a highly obscured AGN with distant reflection washing out rapid and strong fluctuations. Restricting the $C_{\rm XIS\ FI}^{\rm BAT}$ to within 20\%\ of unity (say) and carrying out the fit to model T yields \nh\,$\approx$\,(2--9)\,$\times$\,10$^{24}$\,cm$^{-2}$. We also note that we have not attempted more complex models such as the decoupled \mytorus\ configuration which allows greater flexibility in setting the line-of-sight and global torus properties independently. The present data quality is not high enough to warrant models with many more degrees of freedom. In particular, it will be very useful to have contiguous broadband coverage without a data gap around 10 keV in order to mitigate instrument cross-calibration uncertainties. These issues can be better investigated with data from \nustar\ \citep{nustar} and \astroh\ \citep{astroh12}. 

The soft band luminosity of the \apec\ component is $L_{0.5-2}$\,=\,1.6\,$\times$\,10$^{40}$\,erg\,s$^{-1}$ (corrected for Galactic absorption). The luminosity of the hot thermal gas in the host galaxy described by \apec\ should be related to the star formation rate (SFR). Using the relation of \citet{mineo12_hotgas}, we find SFR$_{\rm X-ray}$\,=\,30$_{-16}^{+37}$\,\Msunpyr\ (1-$\sigma$ scatter). This can be compared with the SFR measured in the far-infrared. \obj\ was detected by the \iras\ mission with a bolometric (8--1000\,\micron) luminosity $L_{8-1000}$\,=\,$5\times 10^{10}$\,\Lsun\ ($2\times 10^{44}$\,erg\,s$^{-1}$), based upon the 12--100\,\micron\ fluxes in the \iras\ Faint Source Catalog v2.0 \citep{iras_fsc} and equations from \citet{sandersmirabel96}. Using the far-infrared luminosity vs. SFR relation by \citet{kennicutt98}, we find SFR$_{\rm IR}$\,=\,9\,\Msunpyr. This is close to the lower 1-$\sigma$ limit implied by SFR$_{\rm X-ray}$ and a star formation contribution to both the infrared and soft X-rays appears viable. However, it should be noted that high spectral resolution and high sensitivity observations have shown AGN photoionisation to dominate the soft X-ray emission of nearby Seyferts \citep[e.g. ][]{sako00, kinkhabwala02, guainazzi07}. A complex of photoionisation lines can mimic a thermal spectrum in low spectral resolution data, and in such a case, the soft X-ray contribution related to star formation could decrease and become consistent with the far-infrared. Such a scenario would also affect the level of the scattered component that dominates over energies of $\approx$\,1--4\,keV, and there is some degeneracy between these components. The best way to distinguish between the various physical models for the soft X-ray emission is through high spectral resolution observations, e.g. using the Soft X-ray Spectrometer which will soon be available onboard \astroh\ \citep{astroh12}.

\subsection{AGN luminosity}

The {\em observed} (absorbed) 2--10 keV AGN luminosity of \obj\ is $L_{2-10}^{\rm obs}$\,=\,1.1\,$\times$\,10$^{41}$\,erg\,s$^{-1}$. The intrinsic luminosity must be higher but is largely unknown, even from indirect diagnostics that are often used as bolometric AGN luminosity indicators. These diagnostics include the \oiii$\lambda$5007\AA\ emission line \citep[e.g. ][]{panessa06} and the hot dust AGN emission \citep[e.g. the 12\,\micron\ nuclear continuum luminosity][]{g09_mirxray}. Absolute optical emission line fluxes have not been published, to our knowledge. In the infrared, one may use data from the \wise\ all sky survey \citep{wise}. The latest AllWISE catalogue tabulates pipeline-measured magnitudes based upon profile-fitting on $\sim$\,6\,arcsec scales. We used the standard zeropoint from \citet{jarrett11} and find a monochromatic 12\,\micron\ luminosity in the $W3$ band (which has a central wavelength closest to 12\,\micron) of $7.2\times 10^{42}$\,erg\,s$^{-1}$. This is to be regarded as an upper limit of the monochromatic AGN power given that there is likely to be at least some non-AGN contamination. In fact, as we discuss later, the \wise\ fluxes are very unlikely to be AGN-dominated. 

\obj\ is also included in the large atlas of high angular resolution mid-infrared (hereafter, mid-IR) imaging compiled by \citet{asmus14} where the nucleus is a non-detection on sub-arcsec scales with a luminosity upper limit $L_{12}$\,$\equiv$\,$\lambda L_{\lambda}$\,(12\,\micron)\,$<$\,$2\times 10^{42}$\,erg\,s$^{-1}$. This flux is an extrapolated estimate based upon a single VLT/VISIR \citep{lagage04} filter observation in the PAH2 band centered on 11.25\,\micron\ in the observed frame. The observation was aborted before completion, resulting in a prominent pattern in the background across the image. We checked that such artifacts should not adversely affect the photometry drastically, though multi-band photometry is required to unambiguously prove this to be the case. There is no mid-IR spectrum of the source available so it is unknown if there is a strong Silicate absorption feature affecting this band. 

In X-rays, the intrinsic AGN flux is simply the integral of $A E^{1-\Gamma}$ over the energy range of interest, with $\Gamma$ being the fitted photon index and $A$ the normalization. However, constraining the intrinsic PL flux level of CTAGN can be subject to great uncertainty given the unknown geometry of the absorber \citep[e.g. ][]{yaqoob12}. Most studies of CTAGN do not quote errors on the intrinsic luminosity, or use the uncertainty on the normalization alone as an error on the luminosity for simplicity, not accounting for the joint variation of $\Gamma$. As a result, the uncertainty on the luminosity may be severely underestimated. 

\begin{figure}
  \begin{center}
    \includegraphics[angle=0,width=8.5cm]{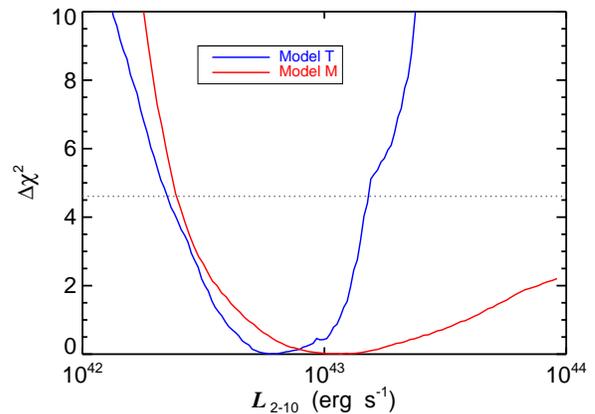}
\caption{The intrinsic PL luminosity uncertainty for models T and M, computed by fitting over a grid of PL photon index and normalization values. The dotted line corresponds to a $\Delta \chi^2$\,=\,4.61 equivalent to a 90\%\ interval for two interesting parameters. 
 \label{fig:lum_confidence}}
  \end{center}
\end{figure}

Here, we measure realistic luminosity uncertainties by computing the joint 90\%\ confidence contours on the two interesting parameters of $A$ and $\Gamma$. The 90\%\ confidence interval on the intrinsic flux (and thus luminosity) is then computed from the flux for all sets of parameters that lie within this 90\%\ confidence region. Fig.\,\ref{fig:lum_confidence} shows the resultant relative $\chi^2$ values for various values of $L_{2-10}$ (i.e. various combinations of $A$ and $\Gamma$). It is clear that there is large uncertainty in $L_{2-10}$. For model T, the 90\% range corresponding to a $\Delta\chi^2$\,=\,4.61 for two interesting parameters spans $\approx 2\times 10^{42}$\,erg\,s$^{-1}$ to $2\times 10^{43}$\,erg\,s$^{-1}$. Model M turns out to be even less constraining with luminosities up to 10$^{44}$\,erg\,s$^{-1}$ allowed. Fixing $\Gamma$ at a value of 1.9 corresponding to the mean observed in well-sampled AGN spectra \citep{mateos05_wide, piconcelli05}, we find $L_{2-10}$\,=\,$4.7_{-1.7}^{+9.4}\times 10^{42}$\,erg\,s$^{-1}$ (model T) and $L_{2-10}$\,=\,$8.7_{-7.0}^{+120}\times 10^{42}$\,erg\,s$^{-1}$ (model M), respectively, with the error in these cases being determined by the normalization uncertainties alone. 

These X-ray luminosities are consistent with known infrared:X-ray luminosity relations for Seyferts \citep[e.g. ][]{horst08, g09_mirxray, asmus11} within the large present uncertainties, though the VLT and \wise\ luminosities quoted above prefer the lower end of the allowed X-ray range. For example, using the $L_{2-10}$ vs. $L_{12}$ relations by \citet{g09_mirxray} with the \wise\ luminosity predicts $L_{2-10}$\,=\,(3--5)\,$\times$\,10$^{42}$\,erg\,s$^{-1}$. Using the VLT/VISIR upper limits instead, the predicted luminosities are $L_{2-10}$\,$<$\,(0.9--2)\,$\times$\,10$^{42}$\,erg\,s$^{-1}$ which are in light tension with the 90\% confidence intervals of $L_{2-10}$ from models M and T. This may be a sign of mild anisotropy in the mid-IR emission of the obscuring torus \citep[e.g. ][]{hoenig11_anisotropy}. Deep multi-band sub-arsec photometry will be able to test this, as will obtaining a mid-IR spectrum to search for any absorption related to the Silicate feature. 

Another potential proxy of the intrinsic luminosity is provided by nuclear water masers. A possible relation between AGN power and masing luminosity has been published by \citet{kondratko06b}, with $L_{2-10}$ being a proxy of the AGN power driving the maser. \citet{braatz96} carried out H$_2$O maser observations of \obj\ and show that the source is undetected down to an isotropic luminosity $L_{\rm H_2O}$\,$\ltsim$\,4\,\Lsun\ (1-$\sigma$). A 3-$\sigma$ limit, together with the \citet{kondratko06b} relation, would predict $L_{2-10}$\,$\ltsim$\,7\,$\times$\,10$^8$\,\Lsun\ (3\,$\times$\,10$^{42}$\,erg\,s$^{-1}$) though a factor of a few higher luminosity is allowed by the relation scatter. 

In summary, our X-ray modelling shows that $L_{2-10}$ lies above $\approx$\,2\,$\times$\,10$^{42}$\,erg\,s$^{-1}$ for both models M and T at 90\%\ confidence. At the upper end, model T lies below $L_{2-10}$\,$\approx$\,2\,$\times$\,10$^{43}$\,erg\,s$^{-1}$, whereas model M is much less constrained and allows values above 10$^{44}$\,erg\,s$^{-1}$. However, two other proxies of the AGN power (the 12\,\micron\ continuum and the water maser power) both prefer the lower end of the allowed $L_{2-10}$ range. As a conservative X-ray estimate, we therefore use the range of uncertainty found from model T, i.e. a 1 dex interval of $L_{2-10}$\,$\approx$\,2\,$\times$\,10$^{42}$--2\,$\times$\,10$^{43}$\,erg\,s$^{-1}$. Measurement of the \oiii\ narrow emission line luminosity will provide an additional indirect diagnostic on the intrinsic power, as will direct sensitive broadband X-ray spectroscopy with \nustar.

\subsection{Comparison to other bona fide CTAGN}

\begin{figure*}
  \begin{center}
    \includegraphics[angle=90,width=12.5cm]{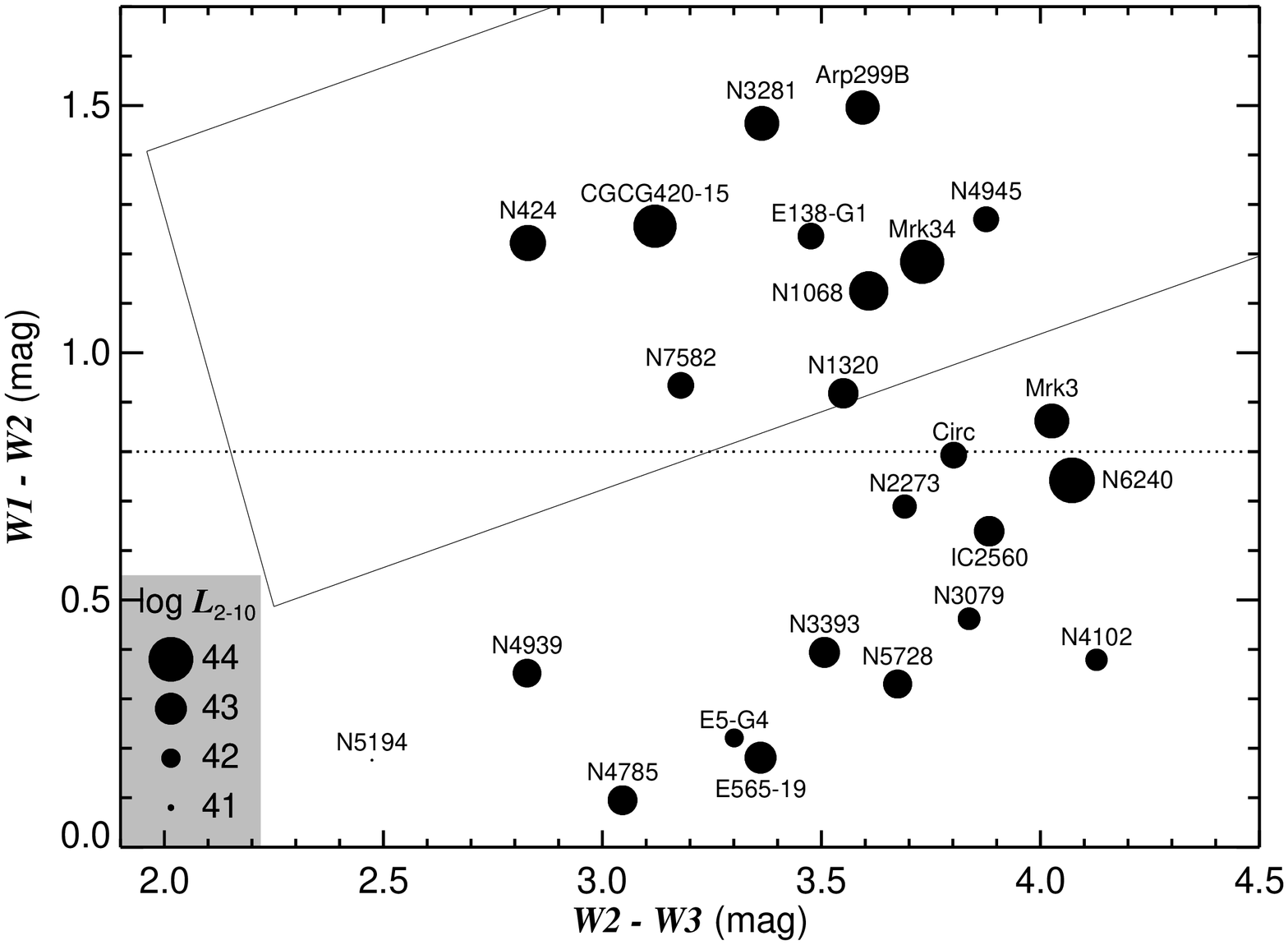}
\caption{\wise\ colour--colour plot for all known bona fide CTAGN from \citet{g14}, now also including \obj. The \lq AGN wedge\rq\ from \citet{mateos12} and the colour cut from \citet{stern12} are shown with the continuous and dashed lines, respectively. Profile-fitting photometry from the AllWISE database is used here, except for Mrk\,3 where the 3-band Post-Cryo \wise\ catalog is used because there is no reported $W3$ flux in the AllWISE database. The sizes of the symbols scale with logarithmic intrinsic 2--10 keV power (size key on the lower left). Median 1-$\sigma$ errors on the colours are approximately 0.03 mag, except for larger errors in the case of two sources with saturated \wise\ photometry NGC\,1068 and Circinus. No correction is made for this minor effect, as this is not relevant for the discussion here. 
 \label{fig:wise_colours}}
  \end{center}
\end{figure*}

How is \obj\ placed in the context of other known bona fide CTAGN? Fig.\,\ref{fig:wise_colours} shows the distribution of known bona fide CTAGN as compiled by \citet[][and references therein]{g14} on the canonical \wise\ colour--colour plane. Increasing levels of AGN contribution to the mid-IR fluxes (relative to the stellar host galaxy contribution) have been shown to move objects approximately upwards in this plot, with most AGN dominated sources found above the colour cut of $W1$\,--\,$W2$\,=\,0.8 proposed by \citet{stern12} and also within the AGN wedge proposed by \citet{mateos12}. It is clear from the plot that the bona fide CTAGN population is quite heterogeneous in its mid-IR colour distribution, with about half of the sources lying below the AGN wedge and cut thresholds. In general, more luminous Seyferts are able to dominate the observed mid-IR fluxes (see also \citealt{asmus14}) and so lie above the thresholds. However, not all sources outside the AGN wedge are intrinsically weak (e.g. NGC\,6240 and Mrk\,3). And of all bona fide CTAGN, \obj\ shows the bluest $W1$\,--\,$W2$ colour, despite the fact the AGN power is not especially weak in X-rays. This could result either from strong host star formation activity contributing to the mid-IR, or from anisotropic and/or weak reprocessed torus emission. Detailed modelling of the broadband spectral energy distribution (SED) will be required to distinguish between these scenarios, which is beyond the scope of this work. The high SFR values derived in Section\,\ref{sec:discussion1} would support an elevated host galaxy stellar dilution to the mid-IR. 

\begin{figure*}
  \begin{center}
    \includegraphics[angle=90,height=10.5cm]{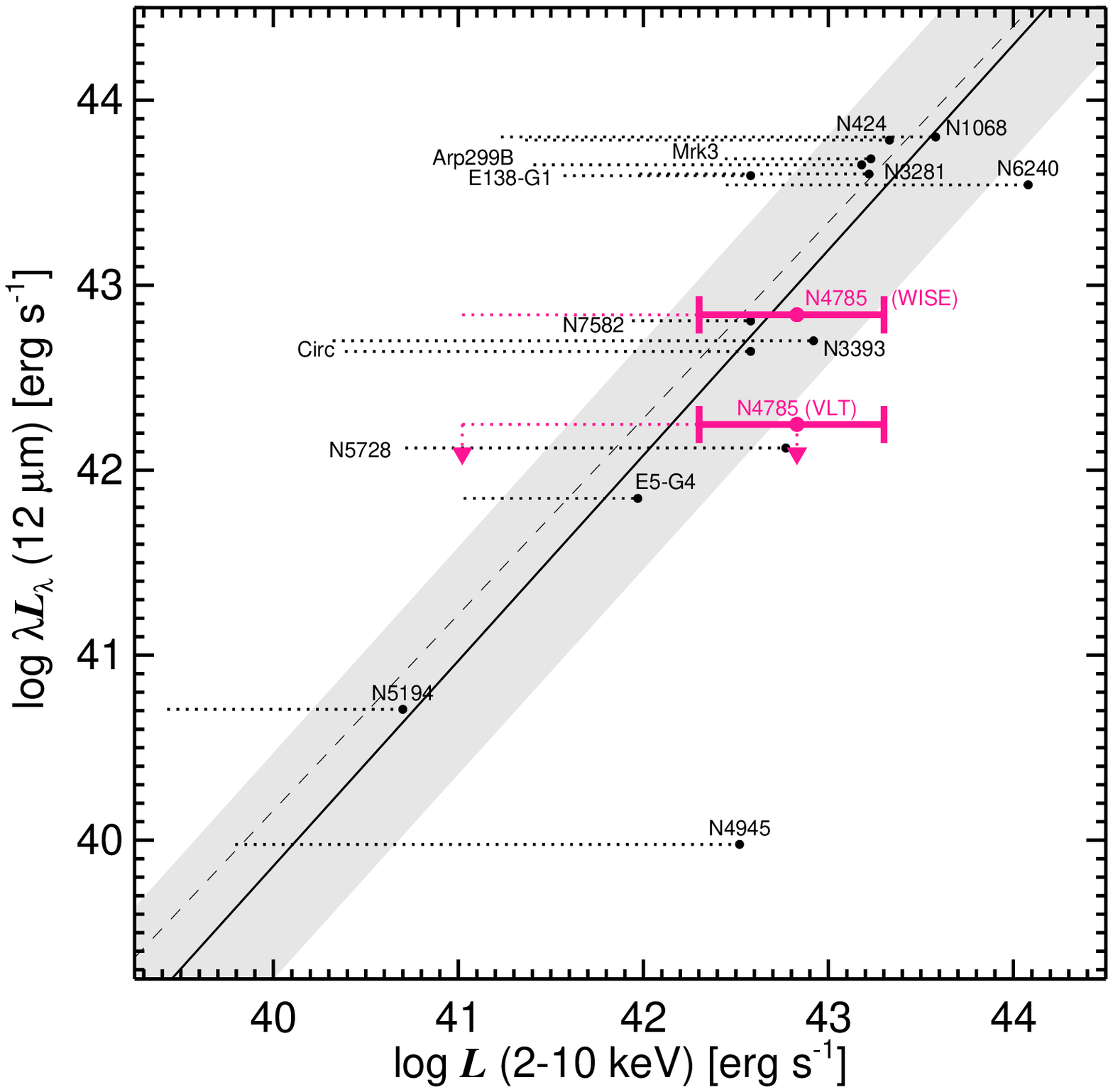}
\caption{Mid-IR vs. X-ray luminosity plot, showing the position of \obj\ in relation to other bona fide CTAGN with sub-arcsec 12\,\micron\ photometry from the high angular resolution mid-IR AGN atlas of \citet{asmus14}. For each source, a dotted horizontal line connects the observed (absorbed) $L_{2-10}$ on the left with the intrinsic (absorption-corrected) power on the right, the latter shown as a filled circle. The X-ray data are compiled in \citet[][]{goulding12}, \citet{g14}, and references therein. The solid line is the mid-IR vs. X-ray relation from \citet{g09_mirxray} with the shaded region being the 2-$\sigma$ relation scatter. For comparison, the dashed line shows a different version of the relation from \citet[][ fitted to include low luminosity AGN]{asmus11}. For \obj, two mid-IR luminosities are plotted: the detection is from \wise\ and the upper-limit denoted by arrows is from the VLT. A horizontal bar is shown in both cases spanning the 90\%\ uncertainty on the intrinsic X-ray luminosity. 
 \label{fig:mirx}}
  \end{center}
\end{figure*}

However, the non-detection of the nucleus in high angular resolution imaging by \citet{asmus14} could instead support an atypical SED for \obj. This is shown in Fig.\,\ref{fig:mirx}, which includes the 15 bona fide CTAGN with mid-IR photometry in the \citet{asmus14} atlas. The absorbed and the intrinsic X-ray luminosities for each source are compared to sub-arcsec 12\,\micron\ photometry. Correction for absorption places almost all sources close to the mid-IR vs. X-ray relation derived for a local sample of Seyfert galaxies by \citet{g09_mirxray}. NGC\,4945 is the strongest known outlier, being underluminous in the mid-IR by about two orders of magnitude for its X-ray power (\citealt{krabbe01, levenson09}; Asmus et al. 2014b, in prep.). In fact, this source is known to be a uniquely prominent outlier in terms of many of its properties, including showing weak and absent coronal lines \citep{goulding09}, potentially possessing a fully embedded AGN \citep{done03} and a putative obscuring torus with atypically low covering factor \citep[e.g. ][]{madejski00, itoh08, puccetti14}. Much of the X-ray reflection signatures usually attributed to a circumnuclear torus instead appear to originate in clouds on much large scales associated with a starburst ring \citep{marinucci12}, which may result in lack of hot dust in the immediate AGN environment. Finally, part of its mid-IR decrement with respect to the mid-IR:X-ray relation may also be related to the edge-on orientation of its host galaxy, with its strong 10\,\micron\ Silicate absorption feature arising on host galaxy scales \citep{goulding12}. 

\obj\ is also underluminous with respect to the mid-IR:X-ray relation, but only mildly based upon the present mid-IR upper-limit (and with the aforementioned caveat about the unknown contribution of Silicate absorption in the high angular resolution data). The inclination angle of the host galaxy is 65.5\,deg according to the HyperLEDA database\footnote{{\tt http://leda.univ-lyon1.fr}}, and there are no obvious signatures of dust lanes in archival optical and near-IR imaging, so this is clearly a different configuration as compared to NGC\,4945. However, the core of \obj\ has been found to possess a complex structure, showing a bright thick double bar as well as an inner disk on scales of $\sim$\,0.3--3\,kpc \citep{marquez99, erwin04}. Mapping the nuclear extinction and cold dust content (e.g. with multi-band optical, infrared and sub-mm observations) will shed more light on the magnitude of large scale obscuration related to the host galaxy, as will obtaining a mid-IR spectrum to search for Silicate absorption. 

In any case, it is clear that if there is a large population of other such CTAGN, they will not be identified easily in the mid-IR alone, even through analyses of the multi-band SEDs, as their AGN mid-IR signatures are sub-dominant. Infrared coronal lines such as the \nev\,$\lambda$\,14.3\,\micron\ line are excellent AGN indicators (\citealt{weaver10}; \citealt{melendez11}; \citealt{goulding12}; Annuar et al. 2014, in prep.), but infrared spectroscopy is highly time-intensive. All-sky X-ray surveys (either with the deepening \swift/BAT survey above 10 keV, or with the upcoming SRG/e-ROSITA mission below 10 keV; \citealt{erosita}) combined with broadband X-ray modelling may be the best way to uncover this population of sources. The samples used to define the already published mid-IR:X-ray relations (e.g. \citealt{krabbe01, lutz04, levenson09, g09_mirxray, asmus11}) do include other edge-on systems, barred and dusty galaxies, and it is not obvious that a significant population of such sources has been missed thus far, but this remains an issue to be investigated with larger complete samples. 

Finally, we note that the presence of Compton thick obscuration has been linked to the presence of stellar host galaxy bars \citep{maiolino99}, and \obj\ appears to support this link. Of the 24 bona fide CTAGN in Fig.\,\ref{fig:wise_colours} with detailed morphological studies, 11 have known bars according to the HyperLEDA database. Another 2 lie in edge on systems, where a bar may be difficult to identify. This suggests that 11 of 22, or 50$_{-18}^{+24}$\%\ of bona fide CTAGN are associated with barred hosts.\footnote{1-$\sigma$ uncertainties on the percentage are determined according to \citealt{gehrels86} and using error propagation.} The overall fraction of spiral galaxies with bars is known to be a strong function of stellar mass \citep{nair10bars}, with a value of $\approx$\,0.25--0.3 (or 25--30\%) for galaxies with a stellar mass of a few times 10$^{10}$\,\Msun\ that should be typical for local hard X-ray detected AGN \citep{koss11}. So there appears to be an excess of CTAGN in barred hosts as compared to the overall galaxy population with bars. However, bar identification has many caveats. For instance, it has been shown that the bar fraction determined from infrared imaging is about twice as high as when using optical imaging \citep{eskridge00}. Therefore, caution is required when interpreting this result, and a detailed assessment of the significance and cause of this suggestive link will be important for understanding the heterogeneity in CTAGN properties.

\section{Acknowledgments}

P.G. thanks STFC for support (grant reference ST/J003697/1). He also acknowledges various discussions with colleagues S.F. H\"{o}nig, D.M. Alexander, A. Annuar and G.B. Lansbury. This research has made use of the Palermo BAT Catalog and database operated at INAF - IASF Palermo. The authors would like to express their thanks to \suzaku\ operations team. This research has made use of the NASA/IPAC Extragalactic Database (NED) which is operated by the Jet Propulsion Laboratory, California Institute of Technology, under contract with the National Aeronautics and Space Administration. \wise\ is a project of Univ. California, Los Angeles, and Jet Propulsion Laboratory (JPL)/California Institute of Technology (Caltech), funded by the NASA. The NASA/IPAC Infrared Science Archive (IRSA) operated by JPL under contract with NASA, was used for querying the infrared databases. The HyperLEDA database was also useful in this work. 

\bibliographystyle{mn2e}
\bibliography{gandhi14b}

\begin{thebibliography}{95}
\expandafter\ifx\csname natexlab\endcsname\relax\def\natexlab#1{#1}\fi

\bibitem[{{Akylas} {et~al}\mbox{.}(2012){Akylas}, {Georgakakis},
  {Georgantopoulos}, {Brightman}, \& {Nandra}}]{akylas12}
{Akylas} A., {Georgakakis} A., {Georgantopoulos} I., {Brightman} M., {Nandra}
  K., 2012, \aap, 546, A98

\bibitem[{{Ar{\'e}valo} {et~al}\mbox{.}(2014){Ar{\'e}valo}, {Bauer},
  {Puccetti}, {Walton}, {Koss}, {Boggs}, {Brandt}, {Brightman}, {Christensen},
  {Comastri}, {Craig}, {Fuerst}, {Gandhi}, {Grefenstette}, {Hailey},
  {Harrison}, {Luo}, {Madejski}, {Madsen}, {Marinucci}, {Matt}, {Saez},
  {Stern}, {Stuhlinger}, {Treister}, {Urry}, \& {Zhang}}]{arevalo14}
{Ar{\'e}valo} P. {et~al.}, 2014, \apj, 791, 81

\bibitem[{{Arnaud}(1996)}]{xspec}
{Arnaud} K.~A., 1996, in ASP Conf. Ser. 101: Astronomical Data Analysis
  Software and Systems V, eds. George H. Jacoby and Jeannette Barnes, Vol.~5,
  pp. 17--+

\bibitem[{{Asmus} {et~al}\mbox{.}(2011){Asmus}, {Gandhi}, {Smette},
  {H{\"o}nig}, \& {Duschl}}]{asmus11}
{Asmus} D., {Gandhi} P., {Smette} A., {H{\"o}nig} S.~F., {Duschl} W.~J., 2011,
  \aap, 536, A36

\bibitem[{{Asmus} {et~al}\mbox{.}(2014){Asmus}, {H{\"o}nig}, {Gandhi},
  {Smette}, \& {Duschl}}]{asmus14}
{Asmus} D., {H{\"o}nig} S.~F., {Gandhi} P., {Smette} A., {Duschl} W.~J., 2014,
  \mnras, 439, 1648

\bibitem[{{Balokovi\'{c} et al.}(2014)}]{balokovic14}
{Balokovi\'{c} et al.}, 2014, ApJ in press

\bibitem[{{Barthelmy} {et~al}\mbox{.}(2005){Barthelmy}, {Barbier}, {Cummings},
  {Fenimore}, {Gehrels}, {Hullinger}, {Krimm}, {Markwardt}, {Palmer},
  {Parsons}, {Sato}, {Suzuki}, {Takahashi}, {Tashiro}, \& {Tueller}}]{bat}
{Barthelmy} S.~D. {et~al.}, 2005, \ssr, 120, 143

\bibitem[{{Baumgartner} {et~al}\mbox{.}(2013){Baumgartner}, {Tueller},
  {Markwardt}, {Skinner}, {Barthelmy}, {Mushotzky}, {Evans}, \&
  {Gehrels}}]{baumgartner13}
{Baumgartner} W.~H., {Tueller} J., {Markwardt} C.~B., {Skinner} G.~K.,
  {Barthelmy} S., {Mushotzky} R.~F., {Evans} P.~A., {Gehrels} N., 2013, \apjs,
  207, 19

\bibitem[{{Blackburn}(1995)}]{ftools}
{Blackburn} J.~K., 1995, in Astronomical Society of the Pacific Conference
  Series, Vol.~77, Astronomical Data Analysis Software and Systems IV,
  {R.~A.~Shaw, H.~E.~Payne, \& J.~J.~E.~Hayes}, ed., pp. 367--+

\bibitem[{{Braatz}, {Wilson} \& {Henkel}(1996){Braatz}, {Wilson}, \&
  {Henkel}}]{braatz96}
{Braatz} J.~A., {Wilson} A.~S., {Henkel} C., 1996, \apjs, 106, 51

\bibitem[{{Brightman} \& {Nandra}(2011)}]{brightmannandra11}
{Brightman} M., {Nandra} K., 2011, \mnras, 413, 1206

\bibitem[{{Brightman} {et~al}\mbox{.}(2014){Brightman}, {Nandra}, {Salvato},
  {Hsu}, \& {Rangel}}]{brightman14}
{Brightman} M., {Nandra} K., {Salvato} M., {Hsu} L.-T., {Rangel} C., 2014,
  ArXiv e-prints

\bibitem[{{Burlon} {et~al}\mbox{.}(2011){Burlon}, {Ajello}, {Greiner},
  {Comastri}, {Merloni}, \& {Gehrels}}]{burlon11}
{Burlon} D., {Ajello} M., {Greiner} J., {Comastri} A., {Merloni} A., {Gehrels}
  N., 2011, \apj, 728, 58

\bibitem[{{Comastri} {et~al}\mbox{.}(1995){Comastri}, {Setti}, {Zamorani}, \&
  {Hasinger}}]{comastri95}
{Comastri} A., {Setti} G., {Zamorani} G., {Hasinger} G., 1995, \aap, 296, 1+

\bibitem[{{Cusumano} {et~al}\mbox{.}(2010{\natexlab{a}}){Cusumano}, {La
  Parola}, {Segreto}, {Ferrigno}, {Maselli}, {Sbarufatti}, {Romano},
  {Chincarini}, {Giommi}, {Masetti}, {Moretti}, {Parisi}, \&
  {Tagliaferri}}]{cusumano10_54}
{Cusumano} G. {et~al.}, 2010{\natexlab{a}}, \aap, 524, A64

\bibitem[{{Cusumano} {et~al}\mbox{.}(2010{\natexlab{b}}){Cusumano}, {La
  Parola}, {Segreto}, {Mangano}, {Ferrigno}, {Maselli}, {Romano}, {Mineo},
  {Sbarufatti}, {Campana}, {Chincarini}, {Giommi}, {Masetti}, {Moretti}, \&
  {Tagliaferri}}]{cusumano10_39}
{Cusumano} G. {et~al.}, 2010{\natexlab{b}}, \aap, 510, A48

\bibitem[{{de Vaucouleurs} {et~al}\mbox{.}(1991){de Vaucouleurs}, {de
  Vaucouleurs}, {Corwin}, {Buta}, {Paturel}, \& {Fouqu{\'e}}}]{rc3}
{de Vaucouleurs} G., {de Vaucouleurs} A., {Corwin}, Jr. H.~G., {Buta} R.~J.,
  {Paturel} G., {Fouqu{\'e}} P., 1991, {Third Reference Catalogue of Bright
  Galaxies. Volume I: Explanations and references. Volume II: Data for galaxies
  between 0$^{h}$ and 12$^{h}$. Volume III: Data for galaxies between 12$^{h}$
  and 24$^{h}$.}

\bibitem[{{Della Ceca} {et~al}\mbox{.}(2008){Della Ceca}, {Severgnini},
  {Caccianiga}, {Comastri}, {Gilli}, {Fiore}, {Piconcelli}, {Malaguti}, \&
  {Vignali}}]{dellaceca08}
{Della Ceca} R. {et~al.}, 2008, \memsai, 79, 65

\bibitem[{{Dickey} \& {Lockman}(1990)}]{dickeylongman90}
{Dickey} J.~M., {Lockman} F.~J., 1990, \araa, 28, 215

\bibitem[{{Done} {et~al}\mbox{.}(2003){Done}, {Madejski}, {{\.Z}ycki}, \&
  {Greenhill}}]{done03}
{Done} C., {Madejski} G.~M., {{\.Z}ycki} P.~T., {Greenhill} L.~J., 2003, \apj,
  588, 763

\bibitem[{{Draper} \& {Ballantyne}(2010)}]{draperballantyne10}
{Draper} A.~R., {Ballantyne} D.~R., 2010, \apjl, 715, L99

\bibitem[{{Erwin}(2004)}]{erwin04}
{Erwin} P., 2004, \aap, 415, 941

\bibitem[{{Eskridge} {et~al}\mbox{.}(2000){Eskridge}, {Frogel}, {Pogge},
  {Quillen}, {Davies}, {DePoy}, {Houdashelt}, {Kuchinski}, {Ram{\'{\i}}rez},
  {Sellgren}, {Terndrup}, \& {Tiede}}]{eskridge00}
{Eskridge} P.~B. {et~al.}, 2000, \aj, 119, 536

\bibitem[{{Fabian} \& {Iwasawa}(1999)}]{fi}
{Fabian} A.~C., {Iwasawa} K., 1999, \mnras, 303, L34

\bibitem[{{Fairall}(1986)}]{fairall86}
{Fairall} A.~P., 1986, \mnras, 218, 453

\bibitem[{{Fukazawa} {et~al}\mbox{.}(2009){Fukazawa}, {Mizuno}, {Watanabe},
  {Kokubun}, {Takahashi}, {Kawano}, {Nishino}, {Sasada}, {Shirai}, {Takahashi},
  {Umeki}, {Yamasaki}, {Yasuda}, {Bamba}, {Ohno}, {Takahashi}, {Ushio},
  {Enoto}, {Kitaguchi}, {Makishima}, {Nakazawa}, {Uehara}, {Yamada}, {Yuasa},
  {Isobe}, {Kawaharada}, {Tanaka}, {Tashiro}, {Terada}, \&
  {Yamaoka}}]{fukazawa09}
{Fukazawa} Y. {et~al.}, 2009, \pasj, 61, 17

\bibitem[{{Gandhi} \& {Fabian}(2003)}]{g03}
{Gandhi} P., {Fabian} A.~C., 2003, \mnras, 339, 1095

\bibitem[{{Gandhi} {et~al}\mbox{.}(2009){Gandhi}, {Horst}, {Smette},
  {H{\"o}nig}, {Comastri}, {Gilli}, {Vignali}, \& {Duschl}}]{g09_mirxray}
{Gandhi} P., {Horst} H., {Smette} A., {H{\"o}nig} S., {Comastri} A., {Gilli}
  R., {Vignali} C., {Duschl} W., 2009, \aap, 502, 457

\bibitem[{{Gandhi} {et~al}\mbox{.}(2014){Gandhi}, {Lansbury}, {Alexander},
  {Stern}, {Ar{\'e}valo}, {Ballantyne}, {Balokovi{\'c}}, {Bauer}, {Boggs},
  {Brandt}, {Brightman}, {Christensen}, {Comastri}, {Craig}, {Del Moro},
  {Elvis}, {Fabian}, {Hailey}, {Harrison}, {Hickox}, {Koss}, {LaMassa}, {Luo},
  {Madejski}, {Ptak}, {Puccetti}, {Teng}, {Urry}, {Walton}, \& {Zhang}}]{g14}
{Gandhi} P. {et~al.}, 2014, ApJ in press, arXiv:1407.1844

\bibitem[{{Gandhi} {et~al}\mbox{.}(2013){Gandhi}, {Terashima}, {Yamada},
  {Mushotzky}, {Ueda}, {Baumgartner}, {Alexander}, {Malzac}, {Vaghmare},
  {Takahashi}, \& {Done}}]{g13_eso565}
{Gandhi} P. {et~al.}, 2013, \apj, 773, 51

\bibitem[{{Gehrels}(1986)}]{gehrels86}
{Gehrels} N., 1986, \apj, 303, 336

\bibitem[{{Gehrels} {et~al}\mbox{.}(2004){Gehrels}, {Chincarini}, {Giommi},
  {Mason}, {Nousek}, {Wells}, {White}, {Barthelmy}, {Burrows}, {Cominsky},
  {Hurley}, {Marshall}, {M{\'e}sz{\'a}ros}, {Roming}, {Angelini}, {Barbier},
  {Belloni}, {Campana}, {Caraveo}, {Chester}, {Citterio}, {Cline}, {Cropper},
  {Cummings}, {Dean}, {Feigelson}, {Fenimore}, {Frail}, {Fruchter}, {Garmire},
  {Gendreau}, {Ghisellini}, {Greiner}, {Hill}, {Hunsberger}, {Krimm},
  {Kulkarni}, {Kumar}, {Lebrun}, {Lloyd-Ronning}, {Markwardt}, {Mattson},
  {Mushotzky}, {Norris}, {Osborne}, {Paczynski}, {Palmer}, {Park}, {Parsons},
  {Paul}, {Rees}, {Reynolds}, {Rhoads}, {Sasseen}, {Schaefer}, {Short},
  {Smale}, {Smith}, {Stella}, {Tagliaferri}, {Takahashi}, {Tashiro},
  {Townsley}, {Tueller}, {Turner}, {Vietri}, {Voges}, {Ward}, {Willingale},
  {Zerbi}, \& {Zhang}}]{swift}
{Gehrels} N. {et~al.}, 2004, \apj, 611, 1005

\bibitem[{{Georgantopoulos} {et~al}\mbox{.}(2013){Georgantopoulos}, {Comastri},
  {Vignali}, {Ranalli}, {Rovilos}, {Iwasawa}, {Gilli}, {Cappelluti}, {Carrera},
  {Fritz}, {Brusa}, {Elbaz}, {Mullaney}, {Castello-Mor}, {Barcons}, {Tozzi},
  {Balestra}, \& {Falocco}}]{georgantopoulos13}
{Georgantopoulos} I. {et~al.}, 2013, \aap, 555, A43

\bibitem[{{Gilli}, {Comastri} \& {Hasinger}(2007){Gilli}, {Comastri}, \&
  {Hasinger}}]{gilli07}
{Gilli} R., {Comastri} A., {Hasinger} G., 2007, \aap, 463, 79

\bibitem[{{Goulding} \& {Alexander}(2009)}]{goulding09}
{Goulding} A.~D., {Alexander} D.~M., 2009, \mnras, 398, 1165

\bibitem[{{Goulding} {et~al}\mbox{.}(2012){Goulding}, {Alexander}, {Bauer},
  {Forman}, {Hickox}, {Jones}, {Mullaney}, \& {Trichas}}]{goulding12}
{Goulding} A.~D., {Alexander} D.~M., {Bauer} F.~E., {Forman} W.~R., {Hickox}
  R.~C., {Jones} C., {Mullaney} J.~R., {Trichas} M., 2012, \apj, 755, 5

\bibitem[{{Guainazzi} \& {Bianchi}(2007)}]{guainazzi07}
{Guainazzi} M., {Bianchi} S., 2007, \mnras, 374, 1290

\bibitem[{{Harrison} {et~al}\mbox{.}(2013){Harrison}, {Craig}, {Christensen},
  {Hailey}, {Zhang}, {Boggs}, {Stern}, {Cook}, {Forster}, {Giommi},
  {Grefenstette}, {Kim}, {Kitaguchi}, {Koglin}, {Madsen}, {Mao}, {Miyasaka},
  {Mori}, {Perri}, {Pivovaroff}, {Puccetti}, {Rana}, {Westergaard}, {Willis},
  {Zoglauer}, {An}, {Bachetti}, {Barri{\`e}re}, {Bellm}, {Bhalerao},
  {Brejnholt}, {Fuerst}, {Liebe}, {Markwardt}, {Nynka}, {Vogel}, {Walton},
  {Wik}, {Alexander}, {Cominsky}, {Hornschemeier}, {Hornstrup}, {Kaspi},
  {Madejski}, {Matt}, {Molendi}, {Smith}, {Tomsick}, {Ajello}, {Ballantyne},
  {Balokovi{\'c}}, {Barret}, {Bauer}, {Blandford}, {Brandt}, {Brenneman},
  {Chiang}, {Chakrabarty}, {Chenevez}, {Comastri}, {Dufour}, {Elvis}, {Fabian},
  {Farrah}, {Fryer}, {Gotthelf}, {Grindlay}, {Helfand}, {Krivonos}, {Meier},
  {Miller}, {Natalucci}, {Ogle}, {Ofek}, {Ptak}, {Reynolds}, {Rigby},
  {Tagliaferri}, {Thorsett}, {Treister}, \& {Urry}}]{nustar}
{Harrison} F.~A. {et~al.}, 2013, \apj, 770, 103

\bibitem[{{H{\"o}nig} {et~al}\mbox{.}(2011){H{\"o}nig}, {Leipski}, {Antonucci},
  \& {Haas}}]{hoenig11_anisotropy}
{H{\"o}nig} S.~F., {Leipski} C., {Antonucci} R., {Haas} M., 2011, \apj, 736, 26

\bibitem[{{Horst} {et~al}\mbox{.}(2008){Horst}, {Gandhi}, {Smette}, \&
  {Duschl}}]{horst08}
{Horst} H., {Gandhi} P., {Smette} A., {Duschl} W.~J., 2008, \aap, 479, 389

\bibitem[{{Ichikawa} {et~al}\mbox{.}(2012){Ichikawa}, {Ueda}, {Terashima},
  {Oyabu}, {Gandhi}, {Matsuta}, \& {Nakagawa}}]{ichikawa12}
{Ichikawa} K., {Ueda} Y., {Terashima} Y., {Oyabu} S., {Gandhi} P., {Matsuta}
  K., {Nakagawa} T., 2012, \apj, 754, 45

\bibitem[{{Itoh} {et~al}\mbox{.}(2008){Itoh}, {Done}, {Makishima}, {Madejski},
  {Awaki}, {Gandhi}, {Isobe}, {Dewangan}, {Griffthis}, {Anabuki}, {Okajima},
  {Reeves}, {Takahashi}, {Ueda}, {Eguchi}, \& {Yaqoob}}]{itoh08}
{Itoh} T. {et~al.}, 2008, \pasj, 60, 251

\bibitem[{{Iwasawa}, {Fabian} \& {Matt}(1997){Iwasawa}, {Fabian}, \&
  {Matt}}]{iwasawa97}
{Iwasawa} K., {Fabian} A.~C., {Matt} G., 1997, \mnras, 289, 443

\bibitem[{{Jarrett} {et~al}\mbox{.}(2011){Jarrett}, {Cohen}, {Masci}, {Wright},
  {Stern}, {Benford}, {Blain}, {Carey}, {Cutri}, {Eisenhardt}, {Lonsdale},
  {Mainzer}, {Marsh}, {Padgett}, {Petty}, {Ressler}, {Skrutskie}, {Stanford},
  {Surace}, {Tsai}, {Wheelock}, \& {Yan}}]{jarrett11}
{Jarrett} T.~H. {et~al.}, 2011, \apj, 735, 112

\bibitem[{{Kennicutt}(1998)}]{kennicutt98}
{Kennicutt}, Jr. R.~C., 1998, \araa, 36, 189

\bibitem[{{Kinkhabwala} {et~al}\mbox{.}(2002){Kinkhabwala}, {Sako}, {Behar},
  {Kahn}, {Paerels}, {Brinkman}, {Kaastra}, {Gu}, \& {Liedahl}}]{kinkhabwala02}
{Kinkhabwala} A. {et~al.}, 2002, \apj, 575, 732

\bibitem[{{Kokubun} {et~al}\mbox{.}(2007){Kokubun}, {Makishima}, {Takahashi},
  {Murakami}, {Tashiro}, {Fukazawa}, {Kamae}, {Madejski}, {Nakazawa},
  {Yamaoka}, {Terada}, {Yonetoku}, {Watanabe}, {Tamagawa}, {Mizuno}, {Kubota},
  {Isobe}, {Takahashi}, {Sato}, {Takahashi}, {Hong}, {Kawaharada}, {Kawano},
  {Mitani}, {Murashima}, {Suzuki}, {Abe}, {Miyawaki}, {Ohno}, {Tanaka},
  {Yanagida}, {Itoh}, {Ohnuki}, {Tamura}, {Endo}, {Hirakuri}, {Hiruta},
  {Kitaguchi}, {Kishishita}, {Sugita}, {Takahashi}, {Takeda}, {Enoto},
  {Hirasawa}, {Katsuta}, {Matsumura}, {Onda}, {Sato}, {Ushio}, {Ishikawa},
  {Murase}, {Odaka}, {Suzuki}, {Yaji}, {Yamada}, {Yamasaki}, {Yuasa}, \& {Hxd
  Team}}]{hxdinorbit}
{Kokubun} M. {et~al.}, 2007, \pasj, 59, 53

\bibitem[{{Kondratko}, {Greenhill} \& {Moran}(2006){Kondratko}, {Greenhill}, \&
  {Moran}}]{kondratko06b}
{Kondratko} P.~T., {Greenhill} L.~J., {Moran} J.~M., 2006, \apj, 652, 136

\bibitem[{{Koss} {et~al}\mbox{.}(2010){Koss}, {Mushotzky}, {Veilleux}, \&
  {Winter}}]{koss10}
{Koss} M., {Mushotzky} R., {Veilleux} S., {Winter} L., 2010, \apjl, 716, L125

\bibitem[{{Koss} {et~al}\mbox{.}(2011){Koss}, {Mushotzky}, {Veilleux},
  {Winter}, {Baumgartner}, {Tueller}, {Gehrels}, \& {Valencic}}]{koss11}
{Koss} M., {Mushotzky} R., {Veilleux} S., {Winter} L.~M., {Baumgartner} W.,
  {Tueller} J., {Gehrels} N., {Valencic} L., 2011, \apj, 739, 57

\bibitem[{{Koyama} {et~al}\mbox{.}(2007){Koyama}, {Tsunemi}, {Dotani}, {Bautz},
  {Hayashida}, {Tsuru}, {Matsumoto}, {Ogawara}, {Ricker}, {Doty}, {Kissel},
  {Foster}, {Nakajima}, {Yamaguchi}, {Mori}, {Sakano}, {Hamaguchi},
  {Nishiuchi}, {Miyata}, {Torii}, {Namiki}, {Katsuda}, {Matsuura}, {Miyauchi},
  {Anabuki}, {Tawa}, {Ozaki}, {Murakami}, {Maeda}, {Ichikawa}, {Prigozhin},
  {Boughan}, {Lamarr}, {Miller}, {Burke}, {Gregory}, {Pillsbury}, {Bamba},
  {Hiraga}, {Senda}, {Katayama}, {Kitamoto}, {Tsujimoto}, {Kohmura}, {Tsuboi},
  \& {Awaki}}]{xis}
{Koyama} K. {et~al.}, 2007, \pasj, 59, 23

\bibitem[{{Krabbe}, {B{\"o}ker} \& {Maiolino}(2001){Krabbe}, {B{\"o}ker}, \&
  {Maiolino}}]{krabbe01}
{Krabbe} A., {B{\"o}ker} T., {Maiolino} R., 2001, \apj, 557, 626

\bibitem[{{Lagage} {et~al}\mbox{.}(2004){Lagage}, {Pel}, {Authier}, {Belorgey},
  {Claret}, {Doucet}, {Dubreuil}, {Durand}, {Elswijk}, {Girardot}, {K{\"a}ufl},
  {Kroes}, {Lortholary}, {Lussignol}, {Marchesi}, {Pantin}, {Peletier},
  {Pirard}, {Pragt}, {Rio}, {Schoenmaker}, {Siebenmorgen}, {Silber}, {Smette},
  {Sterzik}, \& {Veyssiere}}]{lagage04}
{Lagage} P.~O. {et~al.}, 2004, The Messenger, 117, 12

\bibitem[{{Levenson} {et~al}\mbox{.}(2009){Levenson}, {Radomski}, {Packham},
  {Mason}, {Schaefer}, \& {Telesco}}]{levenson09}
{Levenson} N.~A., {Radomski} J.~T., {Packham} C., {Mason} R.~E., {Schaefer}
  J.~J., {Telesco} C.~M., 2009, \apj, 703, 390

\bibitem[{{Lutz} {et~al}\mbox{.}(2004){Lutz}, {Maiolino}, {Spoon}, \&
  {Moorwood}}]{lutz04}
{Lutz} D., {Maiolino} R., {Spoon} H.~W.~W., {Moorwood} A.~F.~M., 2004, \aap,
  418, 465

\bibitem[{{Madejski} {et~al}\mbox{.}(2000){Madejski}, {{\.Z}ycki}, {Done},
  {Valinia}, {Blanco}, {Rothschild}, \& {Turek}}]{madejski00}
{Madejski} G., {{\.Z}ycki} P., {Done} C., {Valinia} A., {Blanco} P.,
  {Rothschild} R., {Turek} B., 2000, \apjl, 535, L87

\bibitem[{{Magdziarz} \& {Zdziarski}(1995)}]{pexrav}
{Magdziarz} P., {Zdziarski} A.~A., 1995, \mnras, 273, 837

\bibitem[{{Maiolino}, {Risaliti} \& {Salvati}(1999){Maiolino}, {Risaliti}, \&
  {Salvati}}]{maiolino99}
{Maiolino} R., {Risaliti} G., {Salvati} M., 1999, \aap, 341, L35

\bibitem[{{Marinucci} {et~al}\mbox{.}(2012){Marinucci}, {Risaliti}, {Wang},
  {Nardini}, {Elvis}, {Fabbiano}, {Bianchi}, \& {Matt}}]{marinucci12}
{Marinucci} A., {Risaliti} G., {Wang} J., {Nardini} E., {Elvis} M., {Fabbiano}
  G., {Bianchi} S., {Matt} G., 2012, \mnras, 423, L6

\bibitem[{{M{\'a}rquez} {et~al}\mbox{.}(1999){M{\'a}rquez}, {Durret},
  {Gonz{\'a}lez Delgado}, {Marrero}, {Masegosa}, {Maza}, {Moles}, {P{\'e}rez},
  \& {Roth}}]{marquez99}
{M{\'a}rquez} I. {et~al.}, 1999, \aaps, 140, 1

\bibitem[{{M{\'a}rquez} {et~al}\mbox{.}(2004){M{\'a}rquez}, {Durret},
  {Masegosa}, {Moles}, {Varela}, {Gonz{\'a}lez Delgado}, {Maza}, {P{\'e}rez},
  \& {Roth}}]{marquez04}
{M{\'a}rquez} I. {et~al.}, 2004, \aap, 416, 475

\bibitem[{{Maselli} {et~al}\mbox{.}(2013){Maselli}, {Massaro}, {Cusumano},
  {D'Abrusco}, {La Parola}, {Paggi}, {Segreto}, {Smith}, \&
  {Tosti}}]{maselli13}
{Maselli} A. {et~al.}, 2013, \apjs, 206, 17

\bibitem[{{Mateos} {et~al}\mbox{.}(2012){Mateos}, {Alonso-Herrero}, {Carrera},
  {Blain}, {Watson}, {Barcons}, {Braito}, {Severgnini}, {Donley}, \&
  {Stern}}]{mateos12}
{Mateos} S. {et~al.}, 2012, \mnras, 426, 3271

\bibitem[{{Mateos} {et~al}\mbox{.}(2005){Mateos}, {Barcons}, {Carrera},
  {Ceballos}, {Caccianiga}, {Lamer}, {Maccacaro}, {Page}, {Schwope}, \&
  {Watson}}]{mateos05_wide}
{Mateos} S. {et~al.}, 2005, \aap, 433, 855

\bibitem[{{Matsuta} {et~al}\mbox{.}(2012){Matsuta}, {Gandhi}, {Dotani},
  {Nakagawa}, {Isobe}, {Ueda}, {Ichikawa}, {Terashima}, {Oyabu}, {Yamamura}, \&
  {Stawarz}}]{matsuta12}
{Matsuta} K. {et~al.}, 2012, \apj, 753, 104

\bibitem[{{Matt} {et~al}\mbox{.}(2000){Matt}, {Fabian}, {Guainazzi}, {Iwasawa},
  {Bassani}, \& {Malaguti}}]{matt00}
{Matt} G., {Fabian} A.~C., {Guainazzi} M., {Iwasawa} K., {Bassani} L.,
  {Malaguti} G., 2000, \mnras, 318, 173

\bibitem[{{Mel{\'e}ndez} {et~al}\mbox{.}(2011){Mel{\'e}ndez}, {Kraemer},
  {Weaver}, \& {Mushotzky}}]{melendez11}
{Mel{\'e}ndez} M., {Kraemer} S.~B., {Weaver} K.~A., {Mushotzky} R.~F., 2011,
  \apj, 738, 6

\bibitem[{{Merloni} {et~al}\mbox{.}(2012){Merloni}, {Predehl}, {Becker},
  {B{\"o}hringer}, {Boller}, {Brunner}, {Brusa}, {Dennerl}, {Freyberg},
  {Friedrich}, {Georgakakis}, {Haberl}, {Hasinger}, {Meidinger}, {Mohr},
  {Nandra}, {Rau}, {Reiprich}, {Robrade}, {Salvato}, {Santangelo}, {Sasaki},
  {Schwope}, {Wilms}, \& {German eROSITA Consortium}}]{erosita}
{Merloni} A. {et~al.}, 2012, eROSITA Science Book: Mapping the Structure of the
  Energetic Universe, arXiv:1209.3114

\bibitem[{{Mineo}, {Gilfanov} \& {Sunyaev}(2012){Mineo}, {Gilfanov}, \&
  {Sunyaev}}]{mineo12_hotgas}
{Mineo} S., {Gilfanov} M., {Sunyaev} R., 2012, \mnras, 426, 1870

\bibitem[{{Moshir} \& {et al.}(1990)}]{iras_fsc}
{Moshir} M., {et al.}, 1990, in IRAS Faint Source Catalogue, version 2.0
  (1990), p.~0

\bibitem[{{Murphy} \& {Yaqoob}(2009)}]{mytorus}
{Murphy} K.~D., {Yaqoob} T., 2009, \mnras, 397, 1549

\bibitem[{{Nair} \& {Abraham}(2010)}]{nair10bars}
{Nair} P.~B., {Abraham} R.~G., 2010, \apjl, 714, L260

\bibitem[{{Panessa} {et~al}\mbox{.}(2006){Panessa}, {Bassani}, {Cappi},
  {Dadina}, {Barcons}, {Carrera}, {Ho}, \& {Iwasawa}}]{panessa06}
{Panessa} F., {Bassani} L., {Cappi} M., {Dadina} M., {Barcons} X., {Carrera}
  F.~J., {Ho} L.~C., {Iwasawa} K., 2006, \aap, 455, 173

\bibitem[{{Piconcelli} {et~al}\mbox{.}(2005){Piconcelli}, {Jimenez-Bail{\'o}n},
  {Guainazzi}, {Schartel}, {Rodr{\'{\i}}guez-Pascual}, \&
  {Santos-Lle{\'o}}}]{piconcelli05}
{Piconcelli} E., {Jimenez-Bail{\'o}n} E., {Guainazzi} M., {Schartel} N.,
  {Rodr{\'{\i}}guez-Pascual} P.~M., {Santos-Lle{\'o}} M., 2005, \aap, 432, 15

\bibitem[{{Planck Collaboration}(2013)}]{planckcosmology}
{Planck Collaboration}, 2013, To appear in A\&A, arXiv:1303.5076

\bibitem[{{Puccetti} {et~al}\mbox{.}(2014){Puccetti}, {Comastri}, {Fiore},
  {Ar{\'e}valo}, {Risaliti}, {Bauer}, {Brandt}, {Stern}, {Harrison},
  {Alexander}, {Boggs}, {Christensen}, {Craig}, {Gandhi}, {Hailey}, {Koss},
  {Lansbury}, {Luo}, {Madejski}, {Matt}, {Walton}, \& {Zhang}}]{puccetti14}
{Puccetti} S. {et~al.}, 2014, ApJ in press, arXiv:1407.3974

\bibitem[{{Sako} {et~al}\mbox{.}(2000){Sako}, {Kahn}, {Paerels}, \&
  {Liedahl}}]{sako00}
{Sako} M., {Kahn} S.~M., {Paerels} F., {Liedahl} D.~A., 2000, \apjl, 543, L115

\bibitem[{{Sanders} \& {Mirabel}(1996)}]{sandersmirabel96}
{Sanders} D.~B., {Mirabel} I.~F., 1996, \araa, 34, 749

\bibitem[{{Segreto} {et~al}\mbox{.}(2010){Segreto}, {Cusumano}, {Ferrigno}, {La
  Parola}, {Mangano}, {Mineo}, \& {Romano}}]{segreto10}
{Segreto} A., {Cusumano} G., {Ferrigno} C., {La Parola} V., {Mangano} V.,
  {Mineo} T., {Romano} P., 2010, \aap, 510, A47

\bibitem[{{Severgnini} {et~al}\mbox{.}(2011){Severgnini}, {Caccianiga}, {Della
  Ceca}, {Braito}, {Vignali}, {La Parola}, \& {Moretti}}]{severgnini11}
{Severgnini} P., {Caccianiga} A., {Della Ceca} R., {Braito} V., {Vignali} C.,
  {La Parola} V., {Moretti} A., 2011, \aap, 525, A38

\bibitem[{{Smith} {et~al}\mbox{.}(2001){Smith}, {Brickhouse}, {Liedahl}, \&
  {Raymond}}]{apec}
{Smith} R.~K., {Brickhouse} N.~S., {Liedahl} D.~A., {Raymond} J.~C., 2001,
  \apjl, 556, L91

\bibitem[{{Stern} {et~al}\mbox{.}(2012){Stern}, {Assef}, {Benford}, {Blain},
  {Cutri}, {Dey}, {Eisenhardt}, {Griffith}, {Jarrett}, {Lake}, {Masci},
  {Petty}, {Stanford}, {Tsai}, {Wright}, {Yan}, {Harrison}, \&
  {Madsen}}]{stern12}
{Stern} D. {et~al.}, 2012, \apj, 753, 30

\bibitem[{{Takahashi} {et~al}\mbox{.}(2007){Takahashi}, {Abe}, {Endo}, {Endo},
  {Ezoe}, {Fukazawa}, {Hamaya}, {Hirakuri}, {Hong}, {Horii}, {Inoue}, {Isobe},
  {Itoh}, {Iyomoto}, {Kamae}, {Kasama}, {Kataoka}, {Kato}, {Kawaharada},
  {Kawano}, {Kawashima}, {Kawasoe}, {Kishishita}, {Kitaguchi}, {Kobayashi},
  {Kokubun}, {Kotoku}, {Kouda}, {Kubota}, {Kuroda}, {Madejski}, {Makishima},
  {Masukawa}, {Matsumoto}, {Mitani}, {Miyawaki}, {Mizuno}, {Mori}, {Mori},
  {Murashima}, {Murakami}, {Nakazawa}, {Niko}, {Nomachi}, {Okada}, {Ohno},
  {Oonuki}, {Ota}, {Ozawa}, {Sato}, {Shinoda}, {Sugiho}, {Suzuki}, {Taguchi},
  {Takahashi}, {Takahashi}, {Takeda}, {Tamura}, {Tamura}, {Tanaka}, {Tanihata},
  {Tashiro}, {Terada}, {Tominaga}, {Uchiyama}, {Watanabe}, {Yamaoka},
  {Yanagida}, \& {Yonetoku}}]{hxd}
{Takahashi} T. {et~al.}, 2007, \pasj, 59, 35

\bibitem[{{Takahashi} {et~al}\mbox{.}(2012){Takahashi}, {Mitsuda}, {Kelley},
  {Aarts}, {Aharonian}, {Akamatsu}, {Akimoto}, {Allen}, {Anabuki}, {Angelini},
  {Arnaud}, {Asai}, {Audard}, {Awaki}, {Azzarello}, {Baluta}, {Bamba}, {Bando},
  {Bautz}, {Blandford}, {Boyce}, {Brown}, {Cackett}, {Chernyakova}, {Coppi},
  {Costantini}, {de Plaa}, {den Herder}, {DiPirro}, {Done}, {Dotani}, {Doty},
  {Ebisawa}, {Eckart}, {Enoto}, {Ezoe}, {Fabian}, {Ferrigno}, {Foster},
  {Fujimoto}, {Fukazawa}, {Funk}, {Furuzawa}, {Galeazzi}, {Gallo}, {Gandhi},
  {Gendreau}, {Gilmore}, {Haas}, {Haba}, {Hamaguchi}, {Hatsukade}, {Hayashi},
  {Hayashida}, {Hiraga}, {Hirose}, {Hornschemeier}, {Hoshino}, {Hughes},
  {Hwang}, {Iizuka}, {Inoue}, {Ishibashi}, {Ishida}, {Ishimura}, {Ishisaki},
  {Ito}, {Iwata}, {Iyomoto}, {Kaastra}, {Kallman}, {Kamae}, {Kataoka},
  {Katsuda}, {Kawahara}, {Kawaharada}, {Kawai}, {Kawasaki}, {Khangaluyan},
  {Kilbourne}, {Kimura}, {Kinugasa}, {Kitamoto}, {Kitayama}, {Kohmura},
  {Kokubun}, {Kosaka}, {Koujelev}, {Koyama}, {Krimm}, {Kubota}, {Kunieda},
  {LaMassa}, {Laurent}, {Lebrun}, {Leutenegger}, {Limousin}, {Loewenstein},
  {Long}, {Lumb}, {Madejski}, {Maeda}, {Makishima}, {Marchand}, {Markevitch},
  {Matsumoto}, {Matsushita}, {McCammon}, {McNamara}, {Miller}, {Miller},
  {Mineshige}, {Minesugi}, {Mitsuishi}, {Miyazawa}, {Mizuno}, {Mori}, {Mori},
  {Mukai}, {Murakami}, {Murakami}, {Mushotzky}, {Nagano}, {Nagino}, {Nakagawa},
  {Nakajima}, {Nakamori}, {Nakazawa}, {Namba}, {Natsukari}, {Nishioka},
  {Nobukawa}, {Nomachi}, {O'Dell}, {Odaka}, {Ogawa}, {Ogawa}, {Ogi}, {Ohashi},
  {Ohno}, {Ohta}, {Okajima}, {Okamoto}, {Okazaki}, {Ota}, {Ozaki}, {Paerels},
  {Paltani}, {Parmar}, {Petre}, {Pohl}, {Porter}, {Ramsey}, {Reis}, {Reynolds},
  {Russell}, {Safi-Harb}, {Sakai}, {Sameshima}, {Sanders}, {Sato}, {Sato},
  {Sato}, {Sato}, {Sawada}, {Serlemitsos}, {Seta}, {Shibano}, {Shida},
  {Shimada}, {Shinozaki}, {Shirron}, {Simionescu}, {Simmons}, {Smith},
  {Sneiderman}, {Soong}, {Stawarz}, {Sugawara}, {Sugita}, {Sugita},
  {Szymkowiak}, {Tajima}, {Takahashi}, {Takeda}, {Takei}, {Tamagawa}, {Tamura},
  {Tamura}, {Tanaka}, {Tanaka}, {Tashiro}, {Tawara}, {Terada}, {Terashima},
  {Tombesi}, {Tomida}, {Tsuboi}, {Tsujimoto}, {Tsunemi}, {Tsuru}, {Uchida},
  {Uchiyama}, {Uchiyama}, {Ueda}, {Ueno}, {Uno}, {Urry}, {Ursino}, {de Vries},
  {Wada}, {Watanabe}, {Werner}, {White}, {Yamada}, {Yamada}, {Yamaguchi},
  {Yamasaki}, {Yamauchi}, {Yamauchi}, {Yatsu}, {Yonetoku}, {Yoshida}, \&
  {Yuasa}}]{astroh12}
{Takahashi} T. {et~al.}, 2012, in Society of Photo-Optical Instrumentation
  Engineers (SPIE) Conference Series, Vol. 8443, Society of Photo-Optical
  Instrumentation Engineers (SPIE) Conference Series

\bibitem[{{Theureau} {et~al}\mbox{.}(2007){Theureau}, {Hanski}, {Coudreau},
  {Hallet}, \& {Martin}}]{theureau07}
{Theureau} G., {Hanski} M.~O., {Coudreau} N., {Hallet} N., {Martin} J.-M.,
  2007, \aap, 465, 71

\bibitem[{{Treister}, {Urry} \& {Virani}(2009){Treister}, {Urry}, \&
  {Virani}}]{treister09}
{Treister} E., {Urry} C.~M., {Virani} S., 2009, \apj, 696, 110

\bibitem[{{Tueller} {et~al}\mbox{.}(2010){Tueller}, {Baumgartner}, {Markwardt},
  {Skinner}, {Mushotzky}, {Ajello}, {Barthelmy}, {Beardmore}, {Brandt},
  {Burrows}, {Chincarini}, {Campana}, {Cummings}, {Cusumano}, {Evans},
  {Fenimore}, {Gehrels}, {Godet}, {Grupe}, {Holland}, {Kennea}, {Krimm},
  {Koss}, {Moretti}, {Mukai}, {Osborne}, {Okajima}, {Pagani}, {Page}, {Palmer},
  {Parsons}, {Schneider}, {Sakamoto}, {Sambruna}, {Sato}, {Stamatikos},
  {Stroh}, {Ukwata}, \& {Winter}}]{tueller10}
{Tueller} J. {et~al.}, 2010, \apjs, 186, 378

\bibitem[{{Tueller} {et~al}\mbox{.}(2008){Tueller}, {Mushotzky}, {Barthelmy},
  {Cannizzo}, {Gehrels}, {Markwardt}, {Skinner}, \& {Winter}}]{tueller08}
{Tueller} J., {Mushotzky} R.~F., {Barthelmy} S., {Cannizzo} J.~K., {Gehrels}
  N., {Markwardt} C.~B., {Skinner} G.~K., {Winter} L.~M., 2008, \apj, 681, 113

\bibitem[{{Ueda} {et~al}\mbox{.}(2014){Ueda}, {Akiyama}, {Hasinger}, {Miyaji},
  \& {Watson}}]{ueda14}
{Ueda} Y., {Akiyama} M., {Hasinger} G., {Miyaji} T., {Watson} M.~G., 2014,
  \apj, 786, 104

\bibitem[{{Ueda} {et~al}\mbox{.}(2007){Ueda}, {Eguchi}, {Terashima},
  {Mushotzky}, {Tueller}, {Markwardt}, {Gehrels}, {Hashimoto}, \&
  {Potter}}]{ueda07}
{Ueda} Y. {et~al.}, 2007, \apjl, 664, L79

\bibitem[{{Vasudevan} {et~al}\mbox{.}(2013){Vasudevan}, {Brandt}, {Mushotzky},
  {Winter}, {Baumgartner}, {Shimizu}, {Schneider}, \&
  {Nousek}}]{vasudevan13_58m}
{Vasudevan} R.~V., {Brandt} W.~N., {Mushotzky} R.~F., {Winter} L.~M.,
  {Baumgartner} W.~H., {Shimizu} T.~T., {Schneider} D.~P., {Nousek} J., 2013,
  \apj, 763, 111

\bibitem[{{Weaver} {et~al}\mbox{.}(2010){Weaver}, {Mel{\'e}ndez}, {Mushotzky},
  {Kraemer}, {Engle}, {Malumuth}, {Tueller}, {Markwardt}, {Berghea}, {Dudik},
  {Winter}, \& {Armus}}]{weaver10}
{Weaver} K.~A. {et~al.}, 2010, \apj, 716, 1151

\bibitem[{{Winter} {et~al}\mbox{.}(2009){Winter}, {Mushotzky}, {Reynolds}, \&
  {Tueller}}]{winter09}
{Winter} L.~M., {Mushotzky} R.~F., {Reynolds} C.~S., {Tueller} J., 2009, \apj,
  690, 1322

\bibitem[{{Wright} {et~al}\mbox{.}(2010){Wright}, {Eisenhardt}, {Mainzer},
  {Ressler}, {Cutri}, {Jarrett}, {Kirkpatrick}, {Padgett}, {McMillan},
  {Skrutskie}, {Stanford}, {Cohen}, {Walker}, {Mather}, {Leisawitz}, {Gautier},
  {McLean}, {Benford}, {Lonsdale}, {Blain}, {Mendez}, {Irace}, {Duval}, {Liu},
  {Royer}, {Heinrichsen}, {Howard}, {Shannon}, {Kendall}, {Walsh}, {Larsen},
  {Cardon}, {Schick}, {Schwalm}, {Abid}, {Fabinsky}, {Naes}, \& {Tsai}}]{wise}
{Wright} E.~L. {et~al.}, 2010, \aj, 140, 1868

\bibitem[{{Yaqoob}(2012)}]{yaqoob12}
{Yaqoob} T., 2012, \mnras, 423, 3360

\end{thebibliography}

\label{lastpage}
\end{document}